\documentclass[fleqn,usenatbib]{mn2e}
\usepackage{bbm}
\usepackage{amsmath}

\usepackage{amssymb}
\usepackage{graphicx}
\def\gsim{\;\lower4pt\hbox{${\buildrel\displaystyle >\over\sim}$}\;}
\def\lsim{\;\lower4pt\hbox{${\buildrel\displaystyle <\over\sim}$}\;}
\def\grls{\;\lower4pt\hbox{${\buildrel\displaystyle >\over <}$}\;}
\title[Self-Similar Behaviours of a Boiling Gas]
{Self-Similar Dynamics of a Relativistically Hot Gas }
\author[Y.-Q. Lou \& Y. Cao]{Yu-Qing Lou$^{1,2,3,}$\thanks{Email:
louyq@tsinghua.edu.cn; lou@oddjob.uchicago.edu}
and Yi Cao$^{1}$\thanks{y-cao04@mails.tsinghua.edu.cn}\\
$^{1}$ Department of Physics and Tsinghua Centre for
Astrophysics (THCA), Tsinghua University, Beijing, 100084, China;\\
$^{2}$ Department of Astronomy and Astrophysics, the University of
Chicago, 5640 South Ells Avenue, Chicago, IL 60637, USA;\\
$^{3}$ National Astronomical Observatories, Chinese Academy of
Sciences, A20, Datun Road, Beijing, 100021, China}
\begin{document}
\maketitle
\begin{abstract}
In the presence of self-gravity, we investigate the self-similar
dynamics of a relativistically hot gas with or without shocks in
astrophysical processes of stellar core collapse, formation of
compact objects, and supernova remnants with central voids. The
model system is taken to be spherically symmetric and the
conservation of specific entropy along streamlines is adopted for
a relativistic hot gas whose energy-momentum relation is expressed
approximately by $\varepsilon=c\mathbbm{p}$ with $\varepsilon$ and
$\mathbbm{p}$ being the energy and momentum of a particle and $c$
being the speed of light. In terms of equation of state, this
leads to a polytropic index $\gamma=4/3$. The conventional
polytropic gas of $P=\kappa\rho^\gamma$, where $P$ is the thermal
pressure, $\rho$ is the mass density, $\gamma$ is the polytropic
index, and $\kappa$ is a global constant, is included in our
theoretical model framework. Two qualitatively different solution
classes arise according to the values of a simple power-law
scaling index $a$, each of which is analyzed separately and
systematically. With explicit conditions, all sonic critical lines
appear straight. We obtain new asymptotic solutions that exist
only for $\gamma=4/3$. Global and asymptotic solutions in various
limits as well as eigensolutions across sonic critical lines are
derived analytically and numerically with or without shocks. By
specific entropy conservation along streamlines, we extend the
analysis of Goldreich \& Weber for a distribution of {\it
variable} specific entropy with time $t$ and radius $r$ and
discuss consequences in the context of a homologous core collapse
prior to supernovae. As an alternative rebound shock model, we
construct an Einstein-de Sitter explosion with shock connections
with various outer flows including a static outer part of a
singular polytropic sphere (SPS). Under the joint action of
thermal pressure and self-gravity, we can also construct
self-similar solutions with central spherical voids with sharp
density variations along their edges.
%of which there is a steep jump in the density profile, can
%form in the centre of the system due to some perturbations.
\end{abstract}

\section{Introduction}

%{\bf Radiation pressure (e.g., Chandrasekhar 19??; Rybicki \&
%Lightman 1979?), trapped neutrino pressure deep in the stellar
%interior, relativistically degenerate materials (e.g.,
%Chandrasekar 1939?), extremely hot stellar materials, processes
%involved in gamma-ray bursts (GRBs) and so forth. Polytropic index
%$\gamma=4/3$ is a highly idealized situation; we further explore
%this special case for a deeper understanding. }
%
%{\bf Should briefly discuss and cite Hunter (1962) somewhere.}
%
%{\bf Provide a better physical justification of this model with
%$\gamma=4/3$. }
%
%{\bf Study the Goldreich \& Weber (1980) and Yahil (1983) papers
%carefully!}
%
%{\bf Directly cite the paper, such as Yahil (1983); do not use
%\citet{b13} which appears weird. }

Radiation pressure (e.g., Chandrasekhar 1939, 1960; Rybicki \&
Lightman 1979), trapped neutrino pressure deep in the stellar
interior of extremely high nuclear density, relativistically
degenerate materials (e.g., Chandrasekhar 1939), extremely hot
interior materials of stars, and processes likely involved in
gamma-ray bursts (GRBs) etc. may be approximated by an equation of
state with a polytropic index of $\gamma=4/3$. Statistical physics
indicates that the state for a stationary radiation field with
particles of no rest mass such as photons is described by a
polytropic relation with an index $\gamma=4/3$. It is proven that
$\gamma=4/3$ is also a very good approximation for relativistically
hot particles with negligible rest mass. Moreover for the stellar
structure, Chandrasekhar (1939) noted that for an infinitesimal
uniform expansion or contraction of a gas sphere, it involves
precisely a polytropic process of an index $\gamma=4/3$. For a
static equilibrium configuration and a presumed
$P=\kappa\rho^{\gamma}$ with a globally constant $\kappa$, the
virial theorem indicates that $\gamma<4/3$ situations are unstable
and $\gamma=4/3$ corresponds to a transition from unstable to stable
configurations as $\gamma$ increases. When $\gamma=4/3$ for a static
equilibrium configuration, the equilibrium condition is referred to
as the Lane-Emden equation with the total enclosed mass $M$ being
independent of the system radius but dependent upon the value of
$\kappa$.

On the other hand, based on the conventional polytropic equation of
state $P=\kappa\rho^{\gamma}$, where $\kappa$ is a global constant
and $\gamma$ varies from $1$ for an isothermal case to $5/3$ for an
adiabatic process of monatomic gas, astrophysicists explored
properties of hydrodynamic behaviours in diverse contexts, such as
star formation, core formation in molecular clouds and supernova
explosions etc. For catching the basic physics and theoretical
simplicity, most works on gravitational stellar core collapses or
stellar explosions were usually carried out under the spherical
symmetry. Hunter (1962) considered the stability of an equilibrium
system and the collapse process based on a polytropic hydrodynamics.
He demonstrated how a dynamical instability during the collapse
leads to a breakup of the spherically symmetric radial inflow of
gas. Shu (1977) constructed the isothermal expansion-wave collapse
solution (EWCS) with a weak discontinuity; and this self-similar
hydrodynamic model has been further developed in the past three
decades, from the isothermal case (e.g., Shu 1977; Lou \& Shen 2004)
to the polytropic case (e.g., Suto \& Silk 1988; Yahil 1983; Lou \&
Wang 2006), as well as to the logotropic case (e.g., Mclaughlin \&
Pudritz 1997). Observationally, spectral lines of CS, H$_2$CO and
other molecules in star forming regions show that the single peak of
each molecular line splits into double peaks with the blue peak
brighter than the red peak, which has been regarded as a supporting
evidence to the Shu model (e.g. Zhou 1992; Walker, Narayanan \& Boss
1994; Myers et al. 1996). It is generally expected that shock waves
are also involved in self-similar collapse or expansion profiles
(e.g., Tsai \& Hsu 1995; Shu et al. 2002; Bian \& Lou 2005).

Note that all above studies were carried out on either the
isothermal case or the $\gamma\neq4/3$ polytropic case. In one case,
the polytropic case of $\gamma=4/3$ is treated as a limiting case
(Yahil 1983). In contrast, Goldreich \& Weber (1980) directly
considered homologous core collapse for a conventional polytropic
gas with $\gamma=4/3$ by making use of the time reversal invariance.
They concluded that when the pressure decreases by a fraction of no
more than about $3\%$ from a static polytrope in equilibrium, a
homologous core collapse would occur in the stellar interior. On the
other hand, numerical simulation of Bethe et al. (1979) indicated a
fractional reduction of pressure by $26\%$ in order to initiate a
core collapse for a supernova explosion. Goldreich \& Weber (1980)
tried to reduce this difference by introducing an inner core of a
progenitor star; Yahil (1983) performed his polytropic analysis and
treated their result as a limit of $\gamma\rightarrow 4/3^+$.

Meanwhile, specific entropy conservation along streamlines does not
necessarily mean a constant specific entropy everywhere at all
times. A more general distribution would be a variable specific
entropy in time and radius (e.g. Cheng 1977, 1978). Fatuzzo et al.
(2004) introduced a self-similar transformation to formulate a more
general problem, which involves a scaling index $a$ and another
index $q$. The more general equation of state appears to be
$P\propto M^q\rho^\gamma$. The $q=0$ case corresponds to the
conventional polytropic gas. In fact, according to this more general
equation of state, the conservation of mass implies the conservation
of specific entropy along streamlines. Nevertheless, Fatuzzo et al.
(2004) mainly focused on the isothermal cases with nonzero inward
flow speeds far away in molecular clouds (Shen \& Lou 2004).

Our consideration is on a more general polytropic gas with
$\gamma=4/3$ with the specific entropy conserved along
streamlines. By a self-similar transformation, we can approach the
resulting nonlinear ordinary differential equations (ODEs)
systematically.
%find that we can straightforwardly deal with
%$\gamma=4/3$ cases with Fatuzzo et al's transformation.
Solution properties depend on the scaling index $a$.
%By the value of $a$, the situation may be totally different.
Given a distribution of variable specific entropy with time and
radius, the result of Goldreich \& Weber (1980) can be
substantially extended.
%If $\kappa$ has an appropriate radial distribution, results can
%have a better consistency with the numerical simulation than GW's.
Meanwhile, many counterparts of previously known solutions in the
isothermal and conventional polytropic cases can also be derived. In
particular, several new asymptotic solutions unique to $\gamma=4/3$
are also obtained. An important and interesting result of our
analysis is that under the joint action of thermal pressure force
and self-gravity, a central spherical void can form and evolve in a
self-similar manner; this is to be compared with the central
spherical void solution of Fillmore \& Goldreich (1984b) which
considered a collection of collisionless particles under
self-gravity in the Einstein-de Sitter expanding universe.

This paper is structured as follows. Nonlinear adiabatic
hydrodynamic equations in spherical symmetry and self-similar
transformation are described in Section 2 for a polytropic gas
with a polytropic index $\gamma=4/3$. The extensions of the
classical analysis of Goldreich \& Weber (1980) are presented in
Section 3 and further discussed for a homologous stellar core
collapse in Section 6.1. We mainly focus on cases of $q=2/3$ for
various solution properties such as the requirement on the scaling
index $a$, the property of scaling invariance, global analytic
solutions, the sonic singular surface, the straight sonic critical
lines, eigensolutions across the sonic critical line, various
asymptotic solutions, and shock jump conditions in Sections 4. We
analyze various semi-complete numerical solutions with or without
shocks and corresponding results in Section 5. Finally, we
conclude and discuss our main results in Section 6. Three
Appendices A, B and C are included at the end for technical
details of derivations and analyses.

\section{Basic Nonlinear Equations and Self-Similar Transformation}

As a theoretical model formulation, dynamical processes outlined in
introduction are governed by the basic nonlinear hydrodynamic
equations under the assumption of spherical symmetry. We naturally
adopt the spherical polar coordinates $(r,\ \theta,\ \phi)$ in the
analysis. The mass conservation is
\begin{eqnarray}
\frac{\partial M}{\partial t}+u\frac{\partial M}{\partial r}=0
\quad\qquad\mbox{  and  }\quad\qquad \frac{\partial M}{\partial
r}=4\pi r^2\rho\ ,\label{eq01}
\end{eqnarray}
where $M(r,t)$ is the enclosed mass within radius $r$ at time $t$,
the mass density $\rho(r,t)$ is a function of $r$ and $t$ and
$u(r,t)$ is the radial flow velocity. The above two relations in
equation (\ref{eq01}) are equivalent to the mass continuity equation
\begin{eqnarray}
\frac{\partial\rho}{\partial
t}+\frac{1}{r^2}\frac{\partial}{\partial r}(r^2\rho u)=0\ .
\end{eqnarray}
The gas motion is governed by the radial momentum equation
\begin{eqnarray}
\frac{\partial u}{\partial t}+u\frac{\partial u}{\partial
r}=-\frac{1}{\rho}\frac{\partial P}{\partial r}-\frac{GM}{r^2}\ ,
\end{eqnarray}
where $P(r,t)$ is the thermal gas pressure and $G=6.67\times
10^{-8}\hbox{ dyne-cm}^2\hbox{ g}^{-2}$ is the gravitational
constant. The Poisson equation relating the mass density $\rho$ and
the gravitational potential $\Phi(r,t)$ is automatically satisfied
with $\partial\Phi/\partial r=GM/r^2$. Finally, the conservation of
specific entropy $s(r,t)$ along streamlines is simply
\begin{eqnarray}
\left(\frac{\partial}{\partial t}+u\frac{\partial}{\partial
r}\right)\left(\frac{P}{\rho^\gamma}\right)=0\ ,\label{eq04}
\end{eqnarray}
where $\gamma$ is the polytropic index. We note that
$P=\kappa\rho^\gamma$ with a constant $\kappa$ is just a
particular case satisfying equation (\ref{eq04}).
%If we denote the local specific entropy by $s(r,t)$, then equation
%(\ref{eq04}) is the continuity equation of specific entropy.
Combining the conservation laws of mass and specific entropy, the
specific entropy $s(r,t)$ can be an arbitrary function $s=s(M)$ of
the enclosed mass $M(r,t)$. The entropy is a statistical quantity
associated with a large number of particles, it appears that in
this situation, the entropy is frozen in particles along
streamlines. By this consideration, we have
\begin{eqnarray}
\frac{ds}{dt}=\frac{ds}{dM}\frac{dM}{dt}=0\
\end{eqnarray}
with
\begin{eqnarray}
\frac{d}{dt}=\frac{\partial}{\partial t}+u\frac{\partial}{\partial
r}
\end{eqnarray}
being the total time derivative along a streamline.

\subsection{Self-Similar Transformation}

In order to solve for self-similar solutions from these nonlinear
partial differential equations, we introduce the following
self-similar transformation to reduce equations $(\ref{eq01})-
(\ref{eq04})$ to nonlinear ODEs, namely
\begin{eqnarray}
x=At^ar\ ,\qquad \rho=\frac{\alpha(x)}{4\pi Gt^2}\ ,\qquad
M=\frac{m(x)}{A^3Gt^{3a+2}}\ ,\nonumber\\
u=\frac{v(x)}{At^{a+1}}\ ,\qquad P=\frac{p(x)}{4\pi
GA^2t^{2(a+2)}}\ ,\qquad\qquad\quad\label{eq07}
\end{eqnarray}
where $a$ is an important scaling index parameter and $A$ is a
dimensional constant coefficient to make the independent variable
$x$ dimensionless. Here, $\alpha(x)$, $m(x)$, $v(x)$, and $p(x)$ are
functions of $x$ only and are referred to as the reduced density,
enclosed mass, velocity and pressure, respectively. Now with
self-similar transformation (\ref{eq07}), equations
$(\ref{eq01})-(\ref{eq04})$ take the form of
\begin{eqnarray}
(ax+v)\frac{dm}{dx}=(3a+2)m\ ,\label{eq1}\\
\frac{dm}{dx}=x^2\alpha\ ,\label{eq2}\\
(ax+v)\frac{dv}{dx}+\frac{1}{\alpha}\frac{dp}{dx}
=-\frac{m}{x^2}+(a+1)v\ ,\label{eq3}\\
(ax+v)\frac{d}{dx}\log\bigg(\frac{p}{\alpha^\gamma}\bigg)
=2(2+a-\gamma)\ ,\label{eq4}\\
(ax+v)\frac{1}{\alpha}\frac{d\alpha}{dx}+\frac{dv}{dx}
=2\left(1-\frac{v}{x}\right)\ \label{eq5}
\end{eqnarray}
(Fatuzzo et al. 2004; Wang \& Lou 2007). Before proceeding, we note
that these equations are invariant under the following time reversal
transformation, namely
\begin{eqnarray}
r\rightarrow r\ ,\qquad t\rightarrow -t\ ,\qquad u\rightarrow -u\ ,
\nonumber\\
\rho\rightarrow\rho\ , \qquad M\rightarrow M\ ,\qquad P\rightarrow
P\ .
\end{eqnarray}
%{\bf You mean $P$ not $p$ here?}
Therefore any solution can also depict its inverse process as long
as this process is reversible (e.g., not involving shocks). For
example, one solution describing a collapse can be also utilized to
describe an expansion process. More importantly, equation
(\ref{eq1}) implies a division of all cases into three classes by
whether or not scaling parameter $a$ is greater than, equal to or
less than $-2/3$; in general, $a$ is required to be negative.

This requirement of a negative $a$ is not obvious by equations
(\ref{eq07})$-$(\ref{eq5}). By the asymptotic solutions
(\ref{static}) and (\ref{asym1}) at large $x$ derived later, it is
necessary to require $a<0$ for convergent solutions at large $x$.

\section{Homologous Core Collapses}

%{\bf And expansion; time reversal for collapses!}
We first analyze the case of $a=-2/3$ precisely which includes the
classical analysis of Goldreich \& Weber (1980). Their model was
applied to a stellar core collapse under self-gravity prior to the
core bouncing in the context of supernova explosions. By equations
(\ref{eq1}) and (\ref{eq2}), we simply have
\begin{eqnarray}
(ax+v)x^2\alpha=0\ .
\end{eqnarray}
The case of $\alpha=0$ everywhere at all time would be a trivial
solution; for nontrivial solution,
%the mass density profile should not be zero everywhere at all time,
the radial flow velocity is thus given by
\begin{eqnarray}
v=-ax=2x/3\ ,
\end{eqnarray}
and then equation (\ref{eq4}) requires $\gamma-a=2$ leading to
$\gamma=4/3$ precisely. Here $v(x)$ represents an expansion
solution, or a core collapse solution with the time reversal
invariance transformation. Meanwhile, equation (\ref{eq5}) becomes
automatically satisfied under this transformation, giving no further
information or constraint. Taking the derivative of equation
(\ref{eq3}) with respect to $x$, we derive
\begin{eqnarray}
\frac{1}{x^2}\frac{d}{dx}\left(\frac{x^2}{\alpha}
\frac{dp}{dx}\right)=-\alpha+\frac{2}{3}\ .\label{eq16}
\end{eqnarray}
Now these equations are not complete yet and a more general
description of specific entropy distribution as a function of $x$ is
allowed. In other words, we already get $P\propto\rho^{4/3}$ but do
not know the proportional coefficient as a function of $(r,\ t)$
which is associated with the enclosed mass $M(r,\ t)$. In fact, this
point can also been seen by directly comparing $P$ and $\rho$ in
self-similar transformation (\ref{eq07}). Physically,
$\log(P/\rho^{4/3})$ is proportional to the specific entropy
$s(r,t)$ in a polytropic gas. Once we know the distribution of
specific entropy as a function of $x$, the self-similar polytropic
flow is then determined.

The first cut is to take a constant specific entropy everywhere at
all times, i.e., $P=\kappa\rho^{4/3}$ with  $\kappa$ being a global
constant. In fact, this is exactly what Goldreich \& Weber (1980)
did. For $A=(4\pi G/\kappa^3)^{1/6}$ in self-similar transformation
(\ref{eq07}), we immediately obtain $p=\alpha^{4/3}$ and a
second-order ODE for $\alpha$ from equation (\ref{eq16}). We may
write $\alpha=f^3$ for the convenience of comparison and the
second-order nonlinear ODE for $f(x)$ with a central condition is
then
\begin{eqnarray}
\left\{
\begin{array}{l}
    \displaystyle\frac{d^2f}{dx^2}+\frac{2}{x}
    \frac{df}{dx}+\frac{f^3}{4}=\frac{1}{6}\ ,\\
    f'(0)=0\ ,\label{eq17}
\end{array}
\right.
\end{eqnarray}
where $f(0)>0$ is related to the central mass density and is an
adjustable parameter up to $f_c$. In numerical integrations, we
may encounter $f(x)=0$ at a finite $x>0$ under certain conditions.
If this is the case, an outer travelling boundary of the flow
system exists. It is fairly straightforward to solve equation
(\ref{eq17}) numerically by the standard Runge-Kutta scheme (e.g.,
Press et al. 1986).
\begin{figure}
\includegraphics[width=0.5\textwidth]{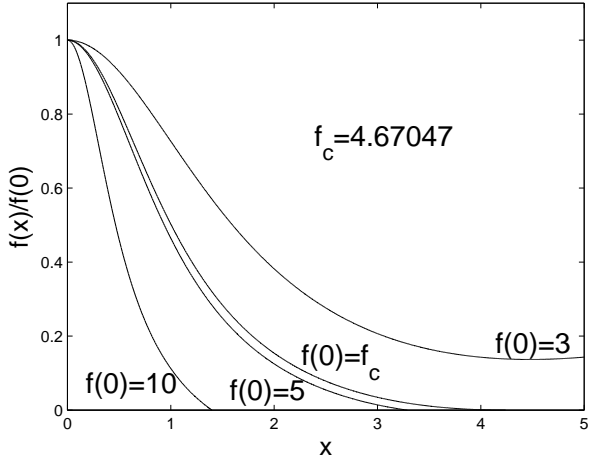}
\caption{Profiles of $f(x)$, closely related to the reduced mass
density by $\alpha(x)=f^3(x)$, with different values of $f(0)$ are
shown for a globally constant specific entropy distribution. Here,
$f(x)$ is normalized by $f(0)$ which is related to the central mass
density. There exists a limiting value $f(0)=f_c$ such that if and
only if $f(0)>f_c$, the solution curve $f(x)$ vanishes at a finite
$x>0$. The value $f_c=4.67047$ is numerically determined. To show
this transition, we take $f(0)=3.0,\ f_c,\ 5.0,\ 10.0$ in turn as
examples of illustration. For $f(0)<f_c$, the solution would have a
density profile of an infinite extent and the radial flow velocity
diverges for large $x$. The curves $f(x)$ give self-similar profiles
of density distribution for a spherical expansion. The process of
collapse can be also described by a time reversal
operation.}\label{GWf}
%{\bf More comments and
%explanations for the last two statements. Collapses and expansions?
%Evaluation of enclosed mass? Shock connection to a gas of different
%$\gamma$? }
\end{figure}

We have just summarized essential results of Goldreich \& Weber
(1980) in their analysis. Through numerical exploration, we also
find that there exists a limiting value for $f(0)$ denoted by
$f_c=4.67047$ (see Figure \ref{GWf}).
%{\bf Can you prove it analytically?}
With $f(0)$ greater than this critical value $f_c$ at $x=0$, the
solution of $f(x)$ is confined by a finite $x$ and has plausible
physical properties. For $f(0)<f_c$, the mass density does not
vanish at a finite $x>0$ and does not approach zero for large $x$
either. By comparing our adjustable parameter $f(0)$ with
parameter $\lambda$ of Goldreich \& Weber (1980), we readily
establish the following simple conversion relation
\begin{eqnarray}
\lambda f^3(0)=2/3\ .
\end{eqnarray}
Parameter $\lambda$ has a maximum value as noted by Goldreich \&
Weber (1980); their maximum value $\lambda_m=0.00654376$ corresponds
to our $f_c$ well. Our $f_c=4.67047$ gives a $\lambda_m=0.00654375$.
%{\bf The value of $f_c$ will be adjusted and thus the relevant numbers.}
%{\bf Please provide $\lambda_c$ and compare specifically.}
When $f(0)$ is lower than this minimum value $f_c$, there is no
self-similar solution with vanishing density at a finite $x\neq 0$.
Physically, $f_c$ corresponds to the minimum central density for a
homologous or self-similar core collapse to be possible.

In addition to the preceding analysis, our polytropic model analysis
here does not necessarily require a constant specific entropy
everywhere (in time and space) and therefore substantially
generalizes the work of Goldreich \& Weber (1980). In fact, we can
allow for a fairly arbitrary distribution of specific entropy and
therefore accommodate a broad class of solutions for the density
profile. A proper distribution of specific entropy can be described
by
\begin{eqnarray}
p=g(x)\alpha^{4/3}\
\end{eqnarray}
where $g(x)$ is a sensible but otherwise arbitrary function. In
fact, the case studied by Goldreich \& Weber (1980) simply
corresponds to $g(x)=1$. For a more general $g(x)$, we readily
derive a second-order nonlinear ODE for $f(x)$ and a central
condition, namely
\begin{eqnarray}
4gf''+\bigg(5g'+\frac{8g}{x}\bigg)f'
+\frac{2}{x}fg'+fg''+f^3=\frac{2}{3}\ ,\nonumber\\
p'(0)=0\Rightarrow g'(0)f(0)+4g(0)f'(0)=0\ ,\label{homo1}
\end{eqnarray}
where prime ``$\prime$" indicates a derivative with respect to $x$
and $\alpha(x)=f^3(x)$ is the reduced mass density. Now given a
value of $f(0)$, related to the central mass density, we can solve
$f(x)$ numerically to determine the self-similar mass density
profile. The intersection of $f(x)$ with the $x$ axis is the moving
`boundary' of the flow system, denoted by $x_b$.

As we know the density profile, we can calculate the enclosed mass
$m(x)$ and the ratio between the central and mean densities. As
shown by equation (\ref{eq2}), the enclosed mass is
\begin{eqnarray}
m(x_b)=\int_0^{x_b}x^2\alpha dx\ .
\end{eqnarray}
%{\bf typo here?} dx\ . added
Using equation (\ref{eq16}), one can readily get
\begin{eqnarray}
m(x_b)=2x_b^3/9-4x_b^2g(x_b)f'(x_b)
\end{eqnarray}
and the dimensional enclosed mass is expressed as
\begin{eqnarray}
M=\frac{m(x_b)}{A^3G}=\frac{1}{A^3G}
\left[\frac{2}{9}x_b^3-4x_b^2g(x_b)f'(x_b)\right]\ .\label{GWM}
\end{eqnarray}
The ratio between the mean and central densities is
\begin{eqnarray}
\frac{\bar{\rho}}{\rho_c}=\frac{3}{f^3(0)}
\left[\frac{2}{9}-\frac{4g(x_b)f'(x_b)}{x_b}\right]\ .
\end{eqnarray}
We are now in a position to make a comparison. With $g(x)=1$,
Goldreich \& Weber (1980; GW) computed the above quantities; within
numerical errors, the results of theirs and ours are mutually
consistent. Our result of $\bar{\rho}/\rho_c$ varies between
$0.0066$ and $0.0185$, while theirs varies between $0.0065$ and
$0.0185$. The value of $m(x_b)$, similar to $r_b^3\bar{\rho}/\rho_c$
in GW, increases by a factor of 1.045 when $f(0)$ increases from
$f_c$ to a sufficiently large value, which is also equal to that of
GW.

As examples of illustration, we shall prescribe specific functional
forms of $g(x)$ and analyze corresponding solutions of $f(x)$
presently in Section 6.3.

\section{Various Solution Properties}

In this section, we mainly focus on cases with $a\neq-2/3$. In
these cases, it is still possible for $\gamma=4/3$ which was not
considered by Goldreich \& Weber (1980) and Yahil (1983).

\subsection{A Preliminary Consideration}

We turn to reduced nonlinear ODEs (\ref{eq1})$-$(\ref{eq5}) to
start our discussion. First, a combination of equations
(\ref{eq1}) and (\ref{eq2}) immediately gives the reduced mass as
\begin{eqnarray}
m(x)=\frac{(ax+v)}{(3a+2)}x^2\alpha(x)\ .\label{meq}
\end{eqnarray}
By equation (\ref{meq}), no confined solution for $\alpha(x)$ by a
finite value of $x>0$ exists because $\alpha=0$ at a finite $x$
directly leads to $m=0$, i.e., no enclosed mass at all within this
$x$ where the mass density vanishes. As a result, no solutions can
be confined by a finite $x>0$. In these cases, the $x$ range of both
analytical and numerical solutions is infinite and some sensible
cutoffs need to be introduced for astrophysical applications.
Moreover, the enclosed mass should be always positive such that
$(ax+v)/(3a+2)>0$. For $a<-2/3$, we must require $v<-ax$, while for
$a>-2/3$, we should have $v>-ax$. This is a strict constraint of
self-similar transformation such that no decreasing solutions of
$v(x)$ exist for $a>-2/3$. In general, $a$ should be always less
than $0$ because of the requirement of a real physical system.
%{\bf Why?}
Dividing equation (\ref{eq4}) by equation
(\ref{eq1}), we obtain
\begin{eqnarray}
p=C_0m^q\alpha^\gamma\ ,\label{fos}
\end{eqnarray}
where another index parameter $q$ is defined by
\begin{eqnarray}
q\equiv 2(2+a-\gamma)/(3a+2)\label{defq}
\end{eqnarray}
and $C_0$ is a constant of integration. This carries an apparent
physical meaning. It is mentioned earlier that if the specific
entropy is a function of the enclosed mass, the conservation of
specific entropy along streamlines is automatically satisfied. The
similarity transformation then gives a more specific constraint on
the form of this function, which is proportional to $M^{q}$.
Substituting the dimensionless quantities for dimensional ones, we
explicitly obtain
\begin{eqnarray}
P=C_0A^{3q-2}(4\pi G)^{\gamma-1}G^qM^q\rho^\gamma\ .\label{state}
\end{eqnarray}
Here we have two constant coefficients: $A$ is introduced in the
transformation and $C_0$ is a constant of integration. For
$\gamma\neq 4/3$ and thus $q\neq 2/3$, we can adjust the value of
$C_0$ and $A$ such that $C_0=1$ (see Lou \& Wang 2007 for more
details). In this paper, however, we focus on the case of
$\gamma=4/3$ and thus $q=2/3$. The constant $A$ no longer plays an
important role because the exponent index vanishes in expression
(\ref{state}) and thus disappears. In contrast, $C_0$ becomes vital
in our case under consideration. On one hand, the local specific
entropy is
\begin{eqnarray}
s=\log\left(\frac{P}{\rho^{4/3}}\right)=\log{C_0}+\frac{1}{3}\log(4\pi
G)+\frac{2}{3}\log(GM)\ .
\end{eqnarray}
%{\bf Please give out an explicit expression of sound speed.}
The value of $C_0$ is related to the specific entropy. On the other
hand, the local polytropic sound speed is
\begin{eqnarray}
c_s=\left(\frac{\partial
P}{\partial\rho}\right)_s^{1/2}=\bigg(\frac{4P}{3\rho}\bigg)^{1/2}
=\left(\frac{4C_0\pi^{2/3} GM^{2/3}\rho^{1/3}}{3}\right)^{1/2}
\end{eqnarray}
which is also related to the value of $C_0$. The value of $C_0$
will affect our equations and thus solutions in a nontrivial
manner.

Substituting equations (\ref{fos}) and (\ref{meq}) into equations
(\ref{eq3}) and (\ref{eq5}), we readily obtain two coupled
nonlinear ODEs
\begin{eqnarray}
(ax+v)\frac{dv}{dx}+\frac{4C_0}{3}x^{4/3}
\left(\frac{ax+v}{3a+2}\right)^{2/3}
\frac{d\alpha}{dx}=\qquad\qquad\nonumber\\
\quad-\frac{ax+v}{3a+2}\alpha+(a+1)v-\frac{2C_0}{3}
\left(\frac{ax+v}{3a+2}\right)^{-1/3}x^{4/3}\alpha\  ,\label{eq11}\\
\frac{dv}{dx}+\frac{(ax+v)}{\alpha}\frac{d\alpha}{dx}
=2\left(1-\frac{v}{x}\right)\ .\qquad\label{eq12}
\end{eqnarray}
%{\bf You may want to put in an appendix of
%$v'=...$ and $\alpha'=...$ with all coefficients
%defined systematically and explicitly. }
Explicit expressions of these two equations for $dv/dx$ and
$d\alpha/dx$ are contained in Appendix \ref{a1}. Our subsequent
analysis is based on these two coupled nonlinear ODEs (\ref{eq11})
and (\ref{eq12}). Before a further discussion, one notes that
besides the time reversal invariance, ODEs (\ref{eq11}) and
(\ref{eq12}) are also invariant under the following scaling
transformation, namely
\begin{eqnarray}
x\rightarrow \eta x\ ,\quad\qquad \alpha\rightarrow\alpha\ ,
\quad\qquad
m\rightarrow \eta^3 m\ ,\nonumber\\
v\rightarrow \eta v\ , \quad\qquad p\rightarrow \eta^2 p\
,\label{scaleinvariance}
\end{eqnarray}
where $\eta$ is an arbitrary positive constant. This scale
invariance only exists when $\gamma=4/3$ or $q=2/3$ and brings us
considerable convenience in theoretical analysis.

\subsection{Global Analytic Solutions}

Previously, two kinds of analytic solutions were found, namely, the
static singular polytropic sphere (SPS) solution and the Einstein-de
Sitter expansion solution in the Newtonian regime (e.g., Wang \& Lou
2007). We confirm that for the current special case of $\gamma=4/3$,
these two solutions still exist with certain modifications and
constraints. For the former, we note that no static SPS solution
exists for $0>a>-2/3$ because of inequality $v>-ax>0$.
%{\bf More explanations?! For $a<-2/3$,}
For $a<-2/3$, we can set $v=0$ in the two ODEs and obtain
\begin{eqnarray}
\alpha=Bx^{2/a}\ ,\label{static}\\
C_0=-\frac{(3a+2)}{2(a+2)}\left(\frac{a}{3a+2}\right)^{4/3}\
,\label{staticcond}
\end{eqnarray}
where $B>0$ is an arbitrary positive coefficient. Unlike previous
polytropic models with $\gamma\neq 4/3$, parameter $C_0$ here is
specifically determined by a chosen $a<-2/3$ in the model.
%while an arbitrary positive integration constant
%$B$ appears in the expression of $\alpha(x)$.
It implies that the system requires a special relationship between
the thermal gas pressure and the combination of
$M^{2/3}\rho^{4/3}$ to keep the system in a radial force balance.

The so-called Einstein-de Sitter solution in the Newtonian
approximation with a constant mass density also exists here for
$\gamma=4/3$. By taking a constant density, we obtain
\begin{eqnarray}
v=\frac{2}{3}x\ ,\qquad
\alpha=\frac{2}{3}\Big(1+2\sqrt[3]{3}C_0\Big)^{-1}\ ,\nonumber\\
m=\frac{2x^3}{9}\Big(1+2\sqrt[3]{3}C_0\Big)^{-1}\ ,\label{eqq}
\end{eqnarray}
%{\bf By the time reversal invariance, this can be a homologous collapse?}
where $C_0>0$ is fairly arbitrary. This solution is independent of
$a$ value as long as $a<-2/3$ and describes a homogeneous expansion
in the Newtonian cosmology. In our case, the constant $\alpha$ is
somewhat different from those of the cases with $q=0$ and
$\gamma\neq 4/3$ (i.e., a conventional polytropic gas with a
constant specific entropy everywhere). We also find that this kind
of solutions exists only in two situations: one is $q=0$ and
$a<-2/3$ (see equations (24) and (25) of Fatuzzo et al. (2004)),
%in \citet{b4}),
while the other is $q=2/3$ also with $a<-2/3$ obtained above.
%{\bf How about discrete values of $q\neq 0$ and $a<-2/3$?}

It is easy to prove that the Einstein-de Sitter solution only
exists in two cases for $q=0$ and $q=2/3$. For $\alpha$ being
constant in ODEs (\ref{eq1})$-$(\ref{eq5}), equation (\ref{eq2})
gives
\begin{eqnarray}
m=x^3\alpha/3\ ,\label{ma}
\end{eqnarray}
where a natural boundary condition is simply $m(0)=0$. Equation
(\ref{eq1}) then leads to $v=2x/3$ which also satisfies equation
(\ref{eq5}). It follows from ODEs (\ref{ma}) and (\ref{eq3}) that
\begin{eqnarray}
\alpha^{\gamma}C_0q(\alpha/3)^{q-1}x^{3q-1} =(2/9-\alpha/3)x\
\end{eqnarray}
with $\alpha$ being a constant. It is clear that for $q=0$, we have
$\alpha=2/3$, while for $q=2/3$, we have a different constant
$\alpha=2/(3+2C_03^{4/3})$ as indicated by solution (\ref{eqq}) at
$\gamma=4/3$ and $a<-2/3$. For $q$ not equal to these two special
values, no Einstein-de Sitter solution is possible. However for
$a=-2/3$ precisely as in Section 3, equation (\ref{eq4}) is of a
$0=0$ form and equation (\ref{eq3}) defines a particular form of
$g(x)$, that is,
\begin{eqnarray}
g(x)=\left(\frac{1}{9f}-\frac{f^2}{6}\right)x^2\ ,
\end{eqnarray}
where $f(x)=\alpha^{1/3}$ is also a constant. By the property of
time reversal invariance, this may also describe a particular
homologous collapse with $\alpha(x)$ being constant and a finite
reference radius.

\subsection{Singular Surface and Sonic Critical Line}

When the determinant of the coefficient matrix of equations
(\ref{eq11}) and (\ref{eq12}) vanishes (see Appendix \ref{a1}),
i.e.,
\begin{eqnarray}
(ax+v)^2=\frac{4C_0}{3}\left(\frac{ax+v}
{3a+2}\right)^{2/3}x^{4/3}\alpha\ ,\label{eqc1}
\end{eqnarray}
where the right-hand side is the polytropic sound speed squared,
%{\bf the RHS is proportional to the polytropic sound speed squared?}
the relevant ODEs (\ref{eq11}) and (\ref{eq12}) become singular and
no finite first derivatives can be obtained (see Appendix A). This
singularity determines a particular surface, referred to as the
sonic singular surface in the three variable space of $x$, $v$ and
$\alpha$. Any solution encountering this sonic singular surface
would diverge except for certain special cases which call for
additional requirements. One possibility is to go across the sonic
critical curve with weak discontinuities (e.g., Whitworth \& Summers
1985 for an isothermal gas) or to jump across the sonic singular
surface with shocks (e.g., Tsai \& Hsu 1985; Shu et al. 2002; Bian
\& Lou 2005; Yu, Lou, Bian \& Wu 2006; Lou \& Gao 2006; Lou \& Wang
2006). Another possibility is to go across the sonic critical curve
smoothly, for which the values of $\alpha$ and $v$ as well as
corresponding derivatives satisfies critical conditions at the
intersection point with the sonic singular surface.

\subsubsection{Determination of the Sonic Critical Line}

A necessary condition for the existence of first derivatives
$\alpha'(x)$ and $v'(x)$ is to require
\begin{eqnarray}
-\frac{ax+v}{3a+2}\alpha+(a+1)v-\frac{2C_0}{3}
\left(\frac{ax+v}{3a+2}\right)^{-1/3}x^{4/3}\alpha\nonumber\\
=2(ax+v)\left(1-\frac{v}{x}\right)\ .\label{eqc2}
\end{eqnarray}
This equation defines a unique curve on the sonic singular surface,
which can be crossed by analytically smooth solutions; this curve is
referred to as the sonic critical curve because it is physically
related to the local sound speed $c_s$.

\begin{figure}
\includegraphics[width=0.5\textwidth]{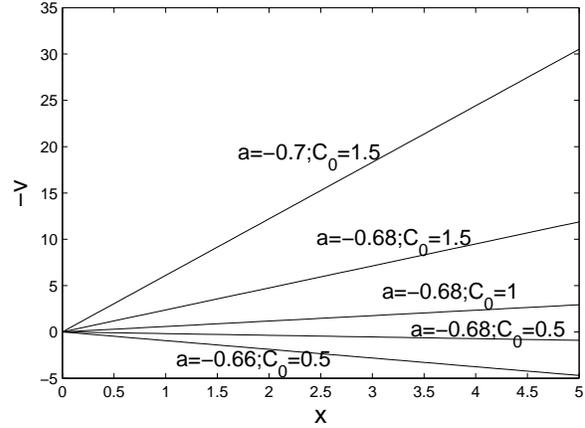}
\caption{Behaviours of the sonic critical lines with different
values of $C_0$ and $a$. From $(a=-0.7,\ C_0=1.5)$ above to
$(a=-0.66,\ C_0=0.5)$ below in order, the corresponding values of
slope $k$ are $-6.1045,\ -2.3754,\ -0.5863,\ 0.1815,\ 0.9399$,
respectively.
%{\bf Please provide corresponding numerical values of $k$.}
They are all rays starting from the origin in the
semi-complete solution space of $-v$ versus $x$.
%{\bf Please redraw the figure with labels
%away from the lines and with larger fonts.}
}\label{Cline}
\end{figure}
Reliable numerical experiments can give us valuable guidance for
conceptual and analytical analysis. Through extensive numerical
exploration (see Figure \ref{Cline}), it is shown that the sonic
critical curve defined as the intersection of two surfaces given by
equations (\ref{eqc1}) and (\ref{eqc2}) seems to be straight lines
starting from the origin in the $-v$ versus $x$ plane and $\alpha$
remains constant along the straight sonic critical lines. Using
equation (\ref{eqc1}) to eliminate $\alpha$ in equation
(\ref{eqc2}), a simple derivation gives $v=kx$ with the slope $k$
being determined by
\begin{eqnarray}
-\frac{3(3a+2)^2}{4C_0}\left(\frac{a+k}{3a+2}\right)^{7/3}+(a+1)k
\qquad\qquad\quad\nonumber\\
\qquad\qquad\quad -(3a+2)(a+k)/2=2(1-k)(a+k)\ .\label{eq13}
\end{eqnarray}
Then, the corresponding value of constant $\alpha$ is given by
\begin{eqnarray}
\alpha=\frac{3(3a+2)^2}{4C_0}\left(\frac{a+k}{3a+2}\right)^{4/3}\
.\label{alpha}
\end{eqnarray}
Once $k$ is solved numerically with given values of parameter pair
$a$ and $C_0$, the relevant sonic critical line is determined with a
corresponding constant $\alpha$. As a consistent confirmation, we
can also use equation (\ref{eqc1}) to eliminate $v$ and then obtain
an algebraic equation for $\alpha$ independent of $x$. The same
conclusion can be reached. Our extensive numerical experiments also
agree with our analytical analysis as expected (see Figure
\ref{Cline}).

For isothermal cases, the projection of the sonic singular surface
coincides with that of the sonic critical line. Thus in the $-v$
versus $x$ plane, the behaviour of critical line can also show that
of the singular surface. Nevertheless, in our situation the shape of
the singular surface is fairly complicated and the critical curve is
just a special curve embedded in it. The projection of the curve in
the $-v$ versus $x$ plane cannot show the exact shape of the entire
singular surface. This is very important in the discussion of shocks
because a shock solution needs to jump across the sonic singular
surface rather than the sonic critical curve.

\subsubsection{Eigensolutions across the Sonic Critical Line}

One solution seldom crosses the sonic singular surface smoothly even
if it meets the sonic critical line. There are also some constraints
on derivatives of proper solutions. For an arbitrary point along the
sonic critical line, denoted by $x_c$ here, we can expand an
analytic solution in terms of Taylor series expansion in the
vicinity of this sonic critical point. Because the nonlinear ODEs is
of second-order, only the first two terms of the series expansion
need to be considered. We write
\begin{eqnarray}
x=x_c+\delta\ ,\qquad v=v_c+\delta v_1\ ,\qquad
\alpha=\alpha_c+\delta\alpha_1\ ,\label{deriv}
\end{eqnarray}
where $v_c$ and $\alpha_c$ are the values at the sonic critical
point with $v_1$ and $\alpha_1$ being the corresponding first
derivatives of $v$ and $\alpha$ at $x_c$, and $\delta$ is a small
displacement away from the critical point $x_c$. Substituting
expression (\ref{deriv}) into coupled nonlinear ODEs (\ref{eq11})
and (\ref{eq12}) and keeping in mind $v_c=kx_c$ and expression
(\ref{alpha}) for $\alpha_c$,
%{\bf You mean $\alpha_c$?},
it is fairly straightforward to derive a quadratic equation for
$v_1$, namely
\begin{eqnarray}
-\frac{7}{3}v_1^2+\left(7+5a-\frac{4}{3}k\right)v_1-\frac{10}{3}k^2+ak+6k
\qquad\quad\nonumber\\
-\frac{3}{2}a^2-6a-6-\frac{\alpha_c\left[a+2(1-k)\right]}{(3a+2)}=0\
.\label{eigen}
\end{eqnarray}
%
%{\bf Can we have EWCS when $k=0$?}
Once the two first derivatives are determined, higher-order
derivatives can be calculated systematically according to nonlinear
ODEs (\ref{eq11}) and (\ref{eq12}). The two eigenvalues of the
velocity first derivative $v_1\equiv dv/dx$ are independent of the
position of critical point $x_c$,
%{\bf You mean $x_c$?},
as shown in equation (\ref{eigen}), which is fundamentally related
to the scaling invariance equation (\ref{scaleinvariance}) discussed
earlier. Generally speaking, quadratic equation (\ref{eigen}) has
two real roots or a pair of complex conjugate roots. Because of the
implicit expression of $k$, we are not able to give an analytical
analysis to decide the existence of real roots. However in our
numerical exploration, all roots are real; that is, for a given
values of $a$ and $C_0$ in our experiments, the eigenvalue problem
has two real roots so far.

Numerical tests also show qualitatively different behaviours
corresponding to the two eigensolutions across the sonic critical
curve. To examine global properties of these two, one can integrate
from the critical point both outwards and inwards, with initial
values given by the series expansion solutions in the vicinity of
the critical point. We call the eigensolution which diverges as $x$
approaching infinity as Type I while the other that converges at
large $x$ is referred to as Type II. The classification of Types I
and II solutions is merely for the convenience of discussion.

\subsection{Various Kinds of Asymptotic Solutions}

In addition to the two global analytic solutions presented above, we
also derive various asymptotic solutions either near the centre
(i.e., $x\rightarrow 0^{+}$) or at infinity (i.e., $x\rightarrow
+\infty$). Because $a>-2/3$ strongly limits solution behaviours such
that $v$ remains always positive and diverges as $x$ approaches
infinity,
%causing that the farther from the center, the larger the
%radial velocity $u$ is, which we do not think is meaningful
%in physics when discussing an infinite solution,
we focus on cases of $a<-2/3$. Previously known asymptotic solutions
will guide us to search for their counterparts in the case of
$\gamma=4/3$ and other possible new solutions should they exist.

By assuming $|v(x)|$ and $\alpha(x)$ to be nonincreasing functions
for large $x$, the first typical asymptotic solutions are
\begin{eqnarray}
\alpha=Hx^{2/a},\quad v=Lx^{(a+1)/a}+Kx^{(2+a)/a}\
,\quad(x\gg1)\label{asym1}
\end{eqnarray}
where $H>0$ and $L$ are two constants of integration, $\gamma=4/3$,
and $K$
%{\bf should be $K$?}
is determined by
\begin{eqnarray}
K=-\frac{aH}{(3a+2)}-2C_0H\left(\frac{a}{3a+2}\right)^{-1/3}
\frac{(a+2)}{(3a+2)}\ .
\end{eqnarray}
%where $\gamma=4/3$ is adopted in the derivation.
The free parameter $L$ was first obtained by
%\citet{b19}
Whitworth \& Summers (1985) for an isothermal gas flow. Cases of
$L=0$ correspond to asymptotic breeze ($K>0$) or contraction ($K<0$)
solutions, depending on whether $v(x)$ is positive or negative at
large $x$. For the first leading term of $v(x)$ in dimensional flow
velocity
\begin{eqnarray}
u\propto Lr^{(a+1)/a}\ ,
\end{eqnarray}
it gives a background flow at infinity and a convergent flow speed
should be required such that $-1\leq a<0$. The sign of $L$ decides
the asymptotic flow direction. Cases of $L>0$ correspond to outflow
or wind solutions while cases of $L<0$ correspond to contraction or
inflow solutions (Lou \& Shen 2004; Lou \& Wang 2006, 2007).
%In general, expression of $K$ involves $\gamma$
%and $q$. For $\gamma<4/3$, the sonic critical curve intersects the
%$x$ axis at a finite $x>0$ and $K$ has a positive lower limit
%corresponding to the limiting solution of the so-called
%expansion-wave collapse solution (EWCS) (Shu 1977). However in our
%model analysis of $\gamma=4/3$, no such positive limit of $K$ exists
%and hence no counterpart of EWCS can be constructed. Another
%perspective of looking at this problem is that the singular
%polytropic sphere (SPS) solution (\ref{static}) and
%(\ref{staticcond}) with $\gamma=4/3$ never encounters the sonic
%critical line except for $x\rightarrow 0^+$ (see Figure 2); there is
%thus no chance for a SPS solution to bifurcate or branch off into a
%solution that eventually matches with the central free-fall
%asymptotic solution at small $x$. {\bf Be careful with the $k=0$
%case for a possible EWCS. }

In addition, another asymptotic solution at large $x$ is described
below; this asymptotic solution may be regarded as a perturbation to
the exact Einstein-de Sitter solution (\ref{eqq}). Assuming a series
expansion solution approaching Einstein-de Sitter solution
(\ref{eqq}) as $x\rightarrow +\infty$, we write
%only with a difference in low order
\begin{eqnarray}
v=\frac{2}{3}x+Ex^{\beta+1}+O\Big(x^{\beta+1}\Big)\ ,
\label{desitterasymV}\qquad\qquad\\
\alpha=\frac{2}{3}\big(1+2\sqrt[3]{3}
C_0\big)^{-1}+Fx^\beta+O\Big(x^\beta\Big)\
,\label{desitterasymA}
\end{eqnarray}
where $E$ and $F$ are two constants to be determined and the $\beta$
parameter is required to be negative, i.e., $\beta<0$. With
$x\rightarrow +\infty$, nonlinear ODEs (\ref{eq11}) and (\ref{eq12})
lead to the following linear homogeneous algebraic equations for $E$
and $F$, namely
\begin{eqnarray}
(\beta+3)\alpha_0E+\bigg(a+\frac{2}{3}\bigg)\beta F=0\ ,\label{neq1}\\
\left[\left(a+\frac{2}{3}\right)\beta
+\frac{\alpha_0}{(3a+2)}\left(1-\frac{2\sqrt[3]{3}}{3}C_0\right)
+\frac{1}{3}\right]E\qquad\quad\nonumber\\
\qquad\qquad\qquad
=-\left(\frac{4\sqrt[3]{3}}{9}C_0\beta+\frac{1}{3}
+\frac{2\sqrt[3]{3}}{3}C_0\right)F\ ,\label{neq2}
\end{eqnarray}
%{\bf should be large ]?}
%{\bf What is $\alpha_0$ here? The Einstein de Sitter constant? }
where $\alpha_0=(2/3)/\big(1+2\sqrt[3]{3}C_0\big)$ is the
Einstein-de Sitter constant reduced density. For nontrivial
solutions of $E$ and $F$, the determinant of the coefficients in
linear equations (\ref{neq1}) and (\ref{neq2}) for $E$ and $F$
should vanish. By this condition, a quadratic equation of $\beta$
appears, namely
\begin{eqnarray}
\Big[(3a+2)^2-4\sqrt[3]{3}C_0\alpha_0\Big]\beta^2
\qquad\qquad\qquad\qquad\nonumber\\
\qquad\qquad\quad +\Big(3a+2-20\sqrt[3]{3}C_0\alpha_0\Big)\beta-6=0\
.\label{quad01}
\end{eqnarray}
After solving quadratic equation (\ref{quad01}), we retain the
negative roots of $\beta$ for the consistency of our approximation.
%smaller than $1$, if existing, for the requirement of the assumption.
We shall show numerical solution examples that approach asymptotic
solutions (\ref{desitterasymV}) and (\ref{desitterasymA}) at large
$x$ presently.
%{\bf What is the requirement on $\beta$?}
%{\bf Have you constructed such a solution example?}

We have just examined asymptotic behaviours of solutions for large
$x$. We now turn to asymptotic solution behaviours in the regime of
small $x$. First, we find a core collapse solution as the
counterpart of the isothermal free-fall solution of Shu (1977),
namely
\begin{eqnarray}
\alpha\simeq|(3a+2)|\bigg[\frac{m(0)}{2}
\bigg]^{1/2}x^{-3/2}\ ,\qquad\qquad\qquad\qquad\nonumber\\
v\simeq \left\{\begin{array}{c}
                -[2m(0)]^{1/2}\ x^{-1/2}\ ,
                 \qquad\mbox{ for }a<-2/3\ ,\\
                 \qquad\ [2m(0)]^{1/2}\ x^{-1/2}\ ,
                 \qquad\mbox{ for }0>a>-2/3\ ,\label{fall}
                 \end{array}\right.
\end{eqnarray}
where $m(0)$ is the limit of $m(x)$ as $x$ goes to $0^{+}$. Because
both mass density and flow velocity are divergent as $x\rightarrow
0^{+}$, a physical cutoff needs to be set somewhere at a small $x$
as the inner reference `boundary' of the flow system under
consideration, or regarded as a reference surface surrounding a
central compact object. Comparing gravity and thermal pressure
force, we immediately have
\begin{eqnarray}
\frac{GM}{r^2}:\frac{1}{\rho}\frac{\partial P}{\partial
r}=\frac{m}{x^2}:\frac{1}{\alpha}\frac{dp}{dx}\sim
x^{-1-3(\gamma-1)/2}\rightarrow\infty\
%,\ x\rightarrow0
\end{eqnarray}
for $x\rightarrow 0^{+}$.
%{\bf second term $\alpha$?} {\bf approach infinity?}
%which diverges rapidly approaching the origin.
Near the centre, the gravity becomes very much stronger than the
thermal pressure force and therefore dominates in the process of
gravitational core collapse. Accreting materials then fall towards
the centre in the form of an almost free fall unimpeded by pressure.
This asymptotic solution represents such a physical scenario that
materials accelerate to fall towards the centre under the
overwhelming self-gravity so that particles gain increasing speed
and acceleration to impact the central object. For black holes,
accreting materials are absorbed more effectively.

As to the so-called Larson-Penston (LP) type solutions at small $x$
(Larson 1969a, b; Penston 1969) with no flow and finite density at
the centre, we can show that the existence of LP type solutions of
$\gamma=4/3$ requires special conditions (Lou \& Shi 2007 in
preparation). With a LP solution in the form of a Taylor series
expansion near the centre, namely
\begin{eqnarray}
v(x)=\sum_{n=0}^{\infty}v_nx^n,\qquad
\alpha(x)=\sum_{n=0}^{\infty}\alpha_nx^n\ ,
\end{eqnarray}
where index $n$ runs through non-negative integers, nonlinear ODEs
(\ref{eq11}) and (\ref{eq12}) require the constant term $v_0$ to be
zero. After straightforward calculations of ODEs by substituting
$v(x)$ and $\alpha(x)$, we just obtain Einstein-de Sitter solution
(\ref{eqq}). Thus solutions with both finite density and velocity
(including the LP type solutions) near the centre may only exist
under rare situations (see Appendix \ref{LP} for details).
%{\bf Not sufficiently clear; difference of $q=0$ and $q\neq 0$!}
%{\bf Could you provide details in an appendix?}

However, if the mass density diverges instead of being finite at
small $x$, a new asymptotic solution can be derived. Let us consider
the leading order terms of such solutions in the form of
\begin{eqnarray}
v=Rx\ ,\qquad \alpha=Nx^\lambda\ , \qquad\mbox{ with }\ \lambda <0\
,\label{new1}
\end{eqnarray}
where $R$, $N$ and $\lambda$ are three parameters to be determined.
Then nonlinear ODEs (\ref{eq11}) and (\ref{eq12}) give
\begin{eqnarray}
\lambda=\frac{2-3R}{a+R}\ ,\qquad\qquad\qquad\qquad\qquad\qquad\\
\frac{2C_0(2+a-2R)}{a+R}+\left(\frac{a+R}{3a+2}\right)^{1/3}=0\
,\qquad\label{calculateR}
\end{eqnarray}
with $N>0$ being arbitrary. The Type II eigensolution mentioned
above just approaches this kind of new asymptotic solution at small
$x$. For $a<-2/3$, we require $R<-a$ in order to keep the enclosed
mass positive and that $\lambda<0$ also requires $R<2/3$. Simple
analysis of equation (\ref{calculateR}) shows that $C_0$ has a
critical value $2^{4/3}/6\approx0.4200$ (see Appendix C for
details), below which $R$ has no real root smaller than $2/3$. For
$C_0\gsim 0.4200$, there two real roots of $R$ for such asymptotic
solutions at small $x$. For $C_0\lsim 0.4200$, this kind of
asymptotic solution does not exist. A possible inference is that
Type II eigensolutions may be truncated before reaching the origin
$x=0$. Later numerical solutions confirm this point.

This solution represents a situation in which the thermal pressure
force is comparable to the self-gravity, that is
\begin{eqnarray}
\frac{GM}{r^2}:\frac{1}{\rho}\frac{\partial P}{\partial
r}=\frac{m}{x^2}:\frac{1}{\alpha}\frac{dp}{dx}\sim
x\alpha^2:x^2\alpha\frac{d\alpha}{dx}\sim1
\end{eqnarray}
As the velocity magnitude decreases at small $x$, the thermal
pressure force actually becomes somewhat larger than the
self-gravity.
%{\bf Is $R$ necessarily negative?} No

Lou \& Wang (2006) found a novel ``quasi-static" solution for a
conventional polytropic gas with $\gamma\neq 4/3$ and proposed a
rebound shock model for supermova explosions (see also Lou \& Wang
2007). We find the counterpart of this ``quasi-static" asymptotic
solution in our case of $\gamma= 4/3$. Static SPS solutions are
described by equation (\ref{static}) and we then introduce
next-order perturbations such that
\begin{eqnarray}
v\simeq Vx^\xi\ ,\qquad\qquad\ \label{quasi1}\\
\alpha \simeq Bx^{2/a}+Wx^\sigma\ ,\label{quasi2}
\end{eqnarray}
where $\xi$ and $\sigma$ are two exponents and $V$ and $W$ are two
coefficients to be determined.
%{\bf Notation $\rho$ may cause confusion. How about $\epsilon$? }
Note that parameter $C_0$ has been specified by equation
(\ref{staticcond}) when discussing static SPS solutions. It is
natural to require $\xi>1$ and $\sigma>2/a$ in reference to SPS
solutions.
%{\bf You want to switch $H$ to $B$?}
Substituting
expressions (\ref{quasi1}) and (\ref{quasi2}) into nonlinear ODEs
(\ref{eq11}) and (\ref{eq12}) with a sufficiently small $x$ and
$C_0$ expression (\ref{staticcond}), we obtain two equations for
coefficients $V$ and $W$, namely
\begin{eqnarray}
\xi-1=\sigma-\frac{2}{a}\ ,\label{cond01}\\
\bigg(\xi+\frac{2}{a}+2\bigg)V+(a\sigma-2)\frac{W}{B}=0\ ,\label{cond02}\\
-\bigg(2+\frac{2}{a}\bigg)V+(a\sigma-2)\frac{W}{B}=0\
.\label{cond03}
\end{eqnarray}
For nontrivial solutions of $V$ and $W$, we must require the
determinant of equations (\ref{cond02}) and (\ref{cond03}) to
vanish. The resulting equation together with condition
(\ref{cond01}) lead to a quadratic equation of $\xi$ (Lou \& Wang
2006). The relevant root of $\xi$ is
\begin{eqnarray}
\xi=-4(1+1/a)\ ,
\end{eqnarray}
%{\bf Vanishing determinant and equation (54)?}
while the other $\xi=2/a$ root of the quadratic equation is
unacceptable. As $\xi>1$ is required, we then have inequality
$-4/5<a<-2/3$. This appears somewhat different in certain aspects of
Lou \& Wang (2006): (i) index $\xi$ is always real (no possibility
for a pair of complex conjugate roots) and only one root is valid
for $\gamma=4/3$; (ii) occasionally, both real roots of this index
in Lou \& Wang (2006) may be valid for $\gamma\neq 4/3$; (iii) this
index may become a pair of complex conjugate roots for $\gamma\neq
4/3$, leading to asymptotic oscillations; and (iv) the allowed range
of $a=-n$ is larger here for $\gamma=4/3$.
%They got a quadratic equation of $\rho$ while in our case
%the other $\rho$ root of the quadratic equation is unacceptable.

\subsection{Shock Jump Conditions}

When a faster flow catches up to a slower one, a shock wave can form
and propagate in stellar winds, molecular clouds and stellar
interiors. A shock occupies a narrow region with discontinuities in
density, pressure, temperature, entropy and flow velocity (e.g.,
Landau \& Lifshitz 1960). In our model framework, we are interested
in self-similar shocks which are ``fixed" in a self-similar profile
(e.g., Sedov 1959). Besides crossing the sonic singular surface
smoothly, a flow solution can also jump across it by shocks which
extends physical solutions with various possibilities. In fact,
shock phenomena are ubiquitous in astrophysical systems. Across a
shock front and in the shock framework of reference, conservations
of mass, momentum and energy hold, and so does the second law of
thermodynamics, the increase of entropy from the upstream side to
the downstream side. In this subsection, subscripts $1$ and $2$
always represent physical quantities of upstream and downstream
sides of a shock, respectively.

In the shock framework of reference, the three conservation laws of
mass, momentum and energy correspond to the following three
equations, namely
\begin{eqnarray}
\rho_1(u_1-u_s)=\rho_2(u_2-u_s)\ ,\label{mass}\\
P_1+\rho_1(u_1-u_s)^2=P_2+\rho_2(u_2-u_s)^2\ ,\label{force}\\
(u_1-u_s)^2/2+w_1=(u_2-u_s)^2/2+w_2\label{energy}\ ,
\end{eqnarray}
where $u_s$ is the shock speed in the laboratory framework and $w$
denotes the heat function defined by
\begin{eqnarray}
w\equiv\frac{\gamma P}{(\gamma-1)\rho}=\frac{4P}{\rho}
\end{eqnarray}
for a polytropic gas with $\gamma=4/3$. We usually introduce
self-similar transformations separately for the upstream and
downstream sides of a shock. However unlike previous work of Lou \&
Wang (2006), the parameter $A$ can be arbitrary for the case of
$\gamma=4/3$ and $q=2/3$ so that the same self-similar
transformation is valid on both sides of a shock.\footnote{In the
analysis of Lou \& Wang (2006), the coefficient $A$ is related to
the local sound speeds, which are different for the upstream and
downstream sides of a shock.} Because shocks are ``fixed" in a
self-similar profile, the scaling parameter $a$ is unchanged across
a shock. After this self-similar transformation, the three
conservation equations (\ref{mass}), (\ref{force}) and
(\ref{energy}) can be reduced to
\begin{eqnarray}
\alpha_1\theta_1=\alpha_2\theta_2\ ,\label{cond1}\\
\frac{p_1}{x_s^2}+\alpha_1\theta_1^2=\frac{p_2}{x_s^2}
+\alpha_2\theta_2^2\ ,\label{cond2}\\
\frac{4p_1}{\alpha_1x_{s}^2}+\frac{\theta_1^2}{2}
=\frac{4p_2}{\alpha_2x_{s}^2}+\frac{\theta_2^2}{2}\ ,\label{cond3}
\end{eqnarray}
where $x_s$ is the shock location (also related to the shock speed),
and $\theta_i$ is defined by $v_i/x_{s}+a$. We now manage to solve
the downstream parameters from the upstream parameters. Using
equations (\ref{cond1}) and (\ref{cond2}) to eliminate $p_2$ and
$\alpha_2$, we obtain a quadratic equation for $\theta_2$. Because
of the invariance for exchanging subscripts $1$ and $2$, we neglect
the trivial solution $\theta_1=\theta_2$ and obtain
\begin{eqnarray}
\theta_2=\frac{8p_1}{7\alpha_1\theta_1x_s^2}+\frac{\theta_1}{7}\
,\label{theta2}
\end{eqnarray}
where $x_s$ is a chosen value for $\gamma=4/3$ with no difference
between the upstream and downstream sides of a shock.\footnote{For
$\gamma\neq 4/3$, $x_s$ is generally different on the two sides of a
shock (e.g., Wang \& Lou 2007).} It is then straightforward to solve
for other downstream quantities
\begin{eqnarray}
\alpha_2=\alpha_1\theta_1/\theta_2\ ,\qquad\qquad\quad\ \\
p_2=p_1+(\alpha_1\theta_1^2-\alpha_2\theta_2^2)x_s^2\ .
\end{eqnarray}
We now know $p_2$ and $\alpha_2$ and $m(x)$ is continuous across a
shock, the downstream coefficient $C_{02}=p_2/(m^{2/3}\rho_2^{4/3})$
is then determined.
%{\bf We simply choose $x_s$?}

Besides the three conservation laws, the second law of
thermodynamics will check whether this solution is physically
appropriate. Here the second law is satisfied as long as the
downstream coefficient $C_{02}$ is greater than the upstream
coefficient $C_{01}$. It is also convenient to introduce the Mach
number here
\begin{eqnarray}
{\cal M}_i^2\equiv\frac{(u_i-u_s)^2}{c_i^2}
=\frac{x_s^2\theta_i^2\alpha_i}{\gamma p_i}\ ,
\end{eqnarray}
where $c_i$ ($i=1, 2$) is the sound speed (upstream, downstream).
Hence equation (\ref{theta2}) can be rewritten as
\begin{eqnarray}
\frac{u_2-u_s}{u_1-u_s} =\frac{\theta_2}{\theta_1}=\frac{6}{7{\cal
M}_1^2}+\frac{1}{7}\ ,
\end{eqnarray}
which is equivalent to
\begin{eqnarray}
{\cal M}_2^2=\frac{2+(\gamma-1){\cal M}_1^2}{2\gamma {\cal
M}_1^2-(\gamma-1)}=\frac{{\cal M}_1^2+6}{8{\cal M}_1^2-1}\
\label{Mach}
\end{eqnarray}
in terms of Mach numbers. The increase of specific entropy requires
that the pressure of the upstream side is lower than that of the
downstream side; this leads to several inequalities below
\begin{eqnarray}
\quad {\cal M}_1^2>1,\qquad\qquad {\cal M}_2^2<1,\qquad\qquad
c_1<c_2\ .
\end{eqnarray}
Qualitatively speaking, the upstream flow is supersonic while the
downstream flow is subsonic.

Reciprocally, we can also calculate quantities of the upstream side
from those of the downstream side following the same derivation
procedure for shock conditions (\ref{cond1})$-$(\ref{cond3}). One
should note that from equation (\ref{Mach}), the physical constraint
on ${\cal M}_2^2$ is $1/8<{\cal M}_2^2<1$ for $\gamma=4/3$.
Therefore when we calculate upstream variables from downstream
variables, the self-similar shock position should be chosen within a
certain sensible range such that ${\cal M}_2^2$ falls within this
specified range. Otherwise, no physical solutions for a self-similar
shock can be constructed, i.e., ${\cal M}_1^2$ becomes less than
unity.

During the construction of numerical shock solutions, we sometimes
specify asymptotic solutions at large $x$ and integrate inward. In
this case, we specify physical variables on the upstream side of a
shock first and then derive physical variables on the downstream
side of a shock. We do not encounter troubles in choosing shock
positions. However, in various occasions, we may need to specify
asymptotic solutions at small $x$ and integrate outward. In such
situations, we specify physical variables on the downstream side of
a shock first and then derive physical variables on the upstream
side of a shock. For these cases, we need to make sure that the
downstream Mach number satisfies the inequality $1/8<{\cal M}_2^2<1$
for our chosen shock positions. Otherwise, upstream variables may
become unphysical as this happens in our numerical exploration.

\section{Numerical Solutions and Results}

We have derived global analytical solutions, various asymptotic
solutions at both large and small $x$, two eigensolutions across
straight sonic critical lines for $\gamma=4/3$, and self-similar
shock conditions. We are now in a position to construct various
global semi-complete solutions numerically by utilizing and matching
these solutions. In reference to the sonic singular surface, we
divide all numerical solutions into three classes: those avoiding
the sonic singular surface, those crossing the straight sonic
critical line smoothly and those with shocks. We construct and
discuss these solutions in order.

\subsection{Solutions not Crossing the Singular Surface}

It is straightforward to construct global numerical solutions
without encountering the sonic singular surface. Starting with the
convergent asymptotic solution (\ref{asym1}) at large $x$ (e.g.,
$x=100$ in our numerical experiments), we integrate back towards the
centre. This procedure works fine unless the solution runs into the
sonic singular surface. The solutions diverge near the centre,
approaching the free-fall asymptotic solution (\ref{fall}) as
$x\rightarrow 0^{+}$. Numerical integrations outwards from the
centre tend to be unstable in the sense that the determination of
the two parameters $H$ and $L$ is fairly sensitive to the value of
$m(0)$. The reason is that there is only one parameter $m(0)$ to be
decided in the inner part while the outer part involves two
parameters $H$ and $L$. In numerical procedures, using two
parameters (e.g., $H$ and $L$ in this case) to decide one parameter
(e.g., $m(0)$ in this case) is stable, while an outward integration
from small $x$ to large $x$, using one parameter to decide two
parameters, tends to be sensitive to the numerical accuracy.
%{\bf Still not clear. Please clarify.}
\begin{figure}
\includegraphics[width=0.5\textwidth]{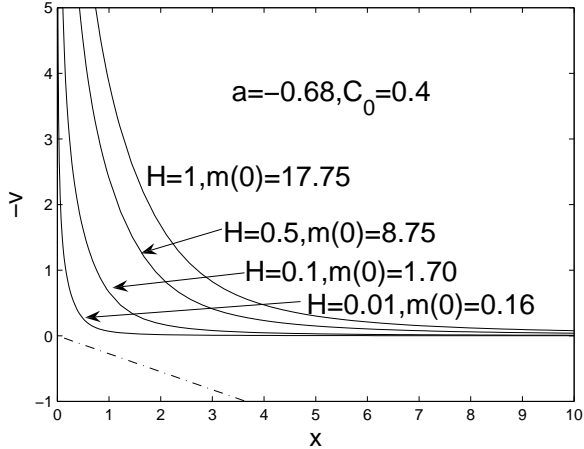}
\caption{Several numerical breeze solutions with $L=0$ are shown as
the solid curves with parameters $H=1,\ 0.5,\ 0.1,\ 0.01$ and
corresponding $m(0)=17.75,\ 8.75,\ 1.70,\ 0.16$ at $a=-0.68$ and
$C_0=0.4$. The straight dashed line passing through the origin is
the sonic critical line. Near the core, all solutions approach the
asymptotic free-fall behaviour (see equation \ref{fall}) at small
$x$.
%{\bf Please try to find the limiting solution by reducing
%$H>0$ and indicate the properties of the two eigensolution.}
%{\bf Please use larger fonts.}
%{\bf Please provide corresponding values of $m(0)$!}
}\label{fig2}
\end{figure}
\begin{figure}
\includegraphics[width=0.5\textwidth]{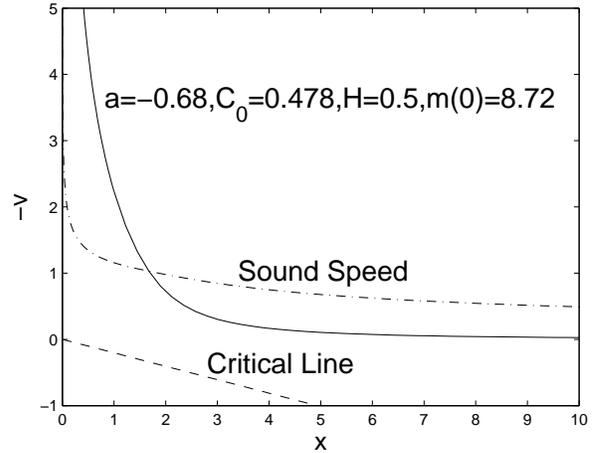}
\caption{A comparison of $-v(x)$ solution with the sound speed
$c(x)$ in the flow. Relevant parameters are: $a=-0.68$, $C_0=0.478$,
$H=0.5$, $m(0)=8.72$. The dashed line is the sonic critical line.
The solid line is the numerical solution of $-v(x)$ while the
dash-dotted line represents the sound speed in the laboratory
framework. The solution $-v(x)$ goes from subsonic at large $x$ to
supersonic at small $x$.}\label{sound}
%successfully exceeds the sound speed without anything
%special happening. {\bf Please use larger fonts. Please
%label sound speed. Provide the value of $m(0)$. } }
\end{figure}
\begin{table}
\begin{center}
\caption{Parameters adopted for examples of numerical solutions are
summarized in this Table 1. All these solutions do not encounter the
sonic critical line. Using the standard Runge-Kutta scheme of fourth
order with initial values calculated from the asymptotic solutions
(\ref{asym1}) at a sufficiently large $x$ (e.g., at $x=100$ in our
numerical integrations). Integrating back towards the origin, we
match the free-fall asymptotic solution (\ref{fall}) as
$x\rightarrow 0^{+}$ and estimate the values of $m(0)$ relevant to
central mass and mass accretion rate.}\label{tb1}
\begin{tabular}{cccccc}
 $a$ & $C_0$ & $H$ & $K$ & $L$ & $m(0)$\\ \hline
 $-0.68$& 0.4&2.0  &$-13.46$&0  &36.0  \\
 $-0.68$& 0.5&2.0  &$-8.33$ &0  &35.8  \\
 $-0.68$& 0.6&2.0  &$-3.20$ &0  &35.6  \\
 $-0.68$& 0.4&1.5  &$-10.99$&0  &26.8  \\
 $-0.68$& 0.4&1.0  &$-6.73$ &0  &17.7  \\
 $-0.68$& 0.4&0.6  &$-4.04$ &0  &10.5  \\
 $-0.68$& 0.4&0.2  &$-1.35$ &0  &3.44  \\
 $-0.68$& 0.4&0.01 &$-0.067$&0  &0.162 \\
 $-0.67$& 0.4&0.001&$-0.041$&0  &0.0661\\
 $-0.70$& 0.4&1.0  &$-1.56$ &0  &7.32  \\
 $-0.72$& 0.4&1.0  &$-0.62$ &0  &4.59  \\
 $-0.74$& 0.4&1.0  &$-0.30$ &0  &3.29  \\
 $-0.67$& 0.4&0.001&$-0.041$&0.2&0.0651\\
 $-0.68$& 0.5&1.0  &$-4.16$ &0.2&17.6  \\
 \hline
\end{tabular}
\end{center}
\end{table}

We adjust the solution parameters $H$ and $L$ and the relevant
parameters $a$ and $C_0$ to explore various solutions (see Fig.
\ref{fig2}). When a solution goes back towards the origin, it
matches with asymptotic free-fall solution (\ref{fall}) and gives
the corresponding $m(0)$ value. For $a=-1$ in expression
(\ref{asym1}), solutions with $L=0$ have vanishing velocities at
infinity; solutions with $L>0$ and $L<0$ correspond to constant
outflows and inflows at infinity, respectively. Such solutions offer
the following scenario: at the beginning time ($t=0$), the gas
system is stationary or has a velocity outwards or inwards with its
mass density profile proportional to $r^{2/a}$.
%{\bf Please clarify! What $L$ are you referring to?}
Under the joint action of self-gravity and thermal pressure force,
the entire system evolves into a central collapse eventually. Around
the central region, the inward self-gravity is always larger than
the thermal pressure force so that materials are accelerated towards
the centre. Nothing singular happens as the local flow speed reaches
the local sound speed (see Fig \ref{sound}). When approaching the
centre, the fluid is almost in a free-fall state. Because of our
presumed spherical symmetry, something must happen around the centre
to destroy the similarity flow or spherical symmetry. For example, a
strong radiation shock may emerge surrounding the centre. Or, a
black hole may take all accreting materials in.
%However, our model can only explain this process. Other models
%are needed to explain or interpret what happens in the center
%to transform this kinetic energy.

\begin{figure}
\includegraphics[width=0.5\textwidth]{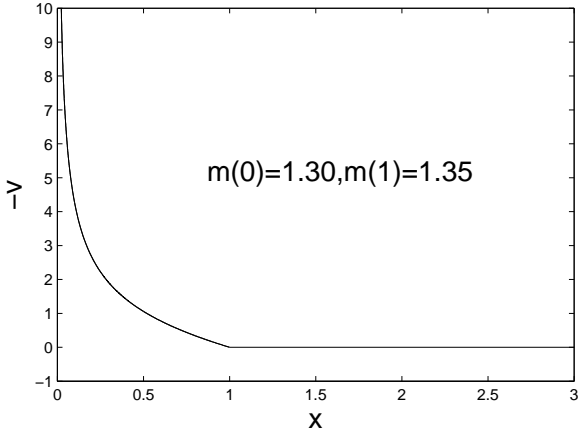}
\caption{The polytropic EWCS with $\gamma=4/3$. The point of weak
discontinuity is $x_e=1$. The outer part of the system is static SPS
while the inner part approaches the free-fall asymptotic solution.
The enclosed mass at the centre $m(0)$ is around $1.30$ and the
collapsing mass $m(1)$ is around $1.35$, indicating that the
majority of materials, around $96.3\%$ concentrates in the centre.
Such polytropic EWCS of $\gamma=4/3$ exists only for $k=0$ with a
pair of parameters $a=-0.68$ and $C_0=0.6623$.
%{\bf Please present $\alpha(x)$ with vertical axis on the right.}
%{\bf Would it be possible to also present $m(x)$?}
}\label{ewcs}
\end{figure}

The isothermal expansion-wave collapse solution (EWCS) of Shu (1977)
was regarded as a limit for a family of solutions without
encountering the sonic critical line. In other words, at a
particular critical point $x_e$ along the sonic critical line, one
of the two eigensolutions leads to the static singular isothermal
sphere (SIS) while the other leads to a solution matching the
central free-fall solution. Thus a semi-complete global solution
with a weak discontinuity is constructed, connecting the two
eigensolutions at $x_e$, with a static SIS for $x>x_e$. Similarly,
for a particular pair of $a$ and $C_0$ values (see equation
\ref{staticcond}), a static SPS with $\gamma=4/3$ exists so that we
can also construct the counterpart of isothermal EWCS. As every
point $x$ is equivalent in the sense of the scaling invariance
(\ref{scaleinvariance}), we simply take $x_e=1$ without loss of
generality.
%{\bf Is $\alpha$ continuous there?}
For this special pair of $a=-0.68$ and $C_0=0.6623$,
%{\bf Please provide specific values}
we have $k=0$ for the slope of the sonic
critical line, and the two corresponding eigensolutions are $v_1=0$
and $v_1=3(1+5a/7)=1.54$.
%{\bf definition of $dv/dx$? the sign is correct?}
Using $v_1=0$ with its corresponding $\alpha_1$ to integrate
outwards, we obtain the outer part of a SPS as the static outer
envelope. Meanwhile, using $v_1=3(1+5a/7)=1.54$
%{\bf sign?}
to integrate back towards the centre, we obtain a central free-fall
solution. Together, we have constructed a polytropic EWCS with
$\gamma=4/3$ (see Fig. \ref{ewcs}).
Let us consider the enclosed mass $m(x)$, where $m(0)$ is the point
mass at the centre and $m(1)-m(0)$ is the mass collapsing towards
the centre. Our numerical result is $m(0)=1.30$ and $m(1)=1.35$.
That is, $96.3\%$ of the total mass concentrates in the central
object and only $3.7\%$ is collapsing towards the centre.

\begin{figure}
\includegraphics[width=0.5\textwidth]{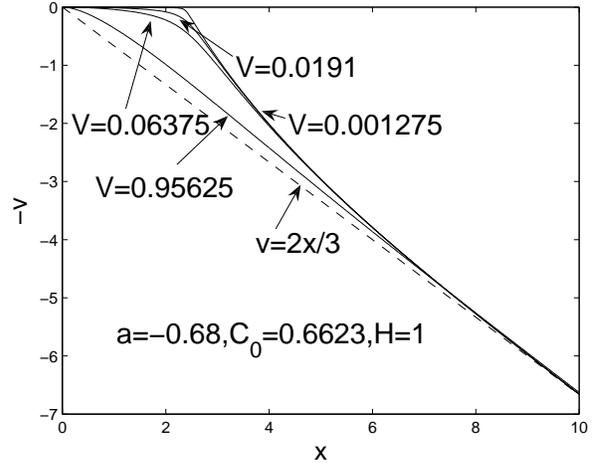}
\caption{%{\bf Please use larger fonts.}
Four quasi-static numerical solutions with $k=0$ and $V>0$ in
quasi-static asymptotic solution (\ref{quasi1}) for small $x$. The
scaling parameter is $a=-0.68$ and the corresponding $C_0=0.6623$ is
computed from equation (\ref{staticcond}).
%{\bf What is the corresponding value of $k$?}
The parameter $B$ in the static solution is set to $1$.
%{\bf Change $H$ to $B$ in the figure.}
By these numerical solutions, it is clear
that the larger the $V$ is, the more rapidly the solution is in the
aberrance of the static state
%{\bf Meaning?}
and approaches the Einstein de Sitter solution
(\ref{desitterasymV}) and (\ref{desitterasymA}).
%These examples show the the instability of the global solution. {\bf Meaning?}
%Even a small perturbation will destruct the balance of the system.
The straight dashed line is $v=2x/3$ for the
Einstein-de
Sitter expansion.}%{\bf Larger fonts!}
\label{quasi_1}
\end{figure}

We can also construct other semi-complete global solutions without
encountering the sonic singular surface. Starting from quasi-static
solution (\ref{quasi1}) and (\ref{quasi2}) at small $x$ with $k=0$
and $V>0$ in equation (\ref{quasi1}), straightforward numerical
integrations lead to global semi-complete solutions without
encountering the sonic singular surface. Figure \ref{quasi_1} shows
such examples at $a=-0.68$ with a corresponding $C_0=0.6623$.
%{\bf What is the corresponding $k$ value for the sonic critical line?}
These solutions never vibrate towards small $x$ according to our
analysis (i.e., no complex conjugate roots are possible). With
increasing $x$, they approach asymptotic expansion solution
(\ref{desitterasymV}) and (\ref{desitterasymA}) rapidly.
%In a certain sense, asymptotic solution (\ref{desitterasymV}) and
%(\ref{desitterasymA}) appears more stable than the static SPS
%solution. {\bf Meaning?}
Numerical results indicate that with a
$V>0$, the gas begins to flow outwards and approaches a constant
density. The larger the value of $V$ is, the stronger the
perturbation is, and the more rapidly the solution approaches the
Einstein-de Sitter expansion phase with $v=2x/3$.

%{\bf Need a clarification for the following paragraph.}
Using new
asymptotic solution (\ref{new1}) at small $x$, we can also construct
global semi-complete solutions. As the sonic critical line is not
enough to describe the relative position between numerical solutions
and the sonic singular surface, we define a velocity $v_c$ such that
\begin{eqnarray}
v_c\equiv-\big(\gamma C_0m^{2/3}\alpha^{1/3}\big)^{1/2}-ax\
,\label{defvc}
\end{eqnarray}
to represent a $v_c$ curve on the singular surface, depending upon
the solution for the reduced enclosed mass $m(x)$ and mass density
$\alpha(x)$ together with adopted parameters $C_0$, $a$ and
$\gamma=4/3$.
%The physical meaning of the defined curve is that: in
%a reference frame which moves outwards with a dimensionless velocity
%$v=-ax$ (in fact, this reference frame is relatively stationary to
%the shock wave which will be discussed next), $v_c$ is the relative
%sound speed because in the definition, the first term is the
%expression of the sound speed in a fixed system and the second term
%represents the relative movement between the fix system and the
%shock.
The purpose is to compare solution $v(x)$ against $v_c$ for the
possibility of encountering the sonic singular surface. From this
definition, it is easy to see that if a solution $v(x)$ meets the
sonic singular surface at some point, this point must the
intersection of the solution curve $v(x)$ and the $v_c$ curve thus
defined. In equation (\ref{new1}), $R,\ N$ and $\lambda$ are three
parameters to be determined. For an appropriate combination of these
parameter values, the solution may not run into the sonic singular
surface and eventually converge to asymptotic solution
(\ref{desitterasymV}) and (\ref{desitterasymA}) at large $x$.
\begin{figure}
\includegraphics[width=0.5\textwidth]{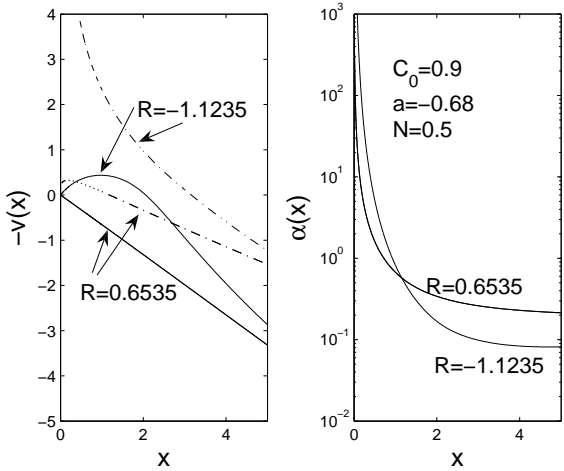}
\caption{
%{\bf Please redraw the figure with relevant corrections
%such $R$ not $M$. One parameter per line. }
An example for the case of $a=-0.68$ and $C_0=0.9$ is illustrated.
The corresponding slope $k=-0.3745$ for the sonic critical line.
It has two real roots of $R$: one is $R=-1.1235$ and the other is
$R=0.6535$. The inner part can be described by asymptotic solution
(\ref{new1}) at small
$x$ with other parameters $N=0.500$ and $\lambda=-2.9778$, %{\bf notations!}
while the outer part can be described by asymptotic solution
(\ref{desitterasymV}) and (\ref{desitterasymA}) at large $x$ with
parameter $\beta=-0.7362$. The two dash-dotted curves in the left
panel are $v_c(x)$ as defined by expression (\ref{defvc}) related to
the relevant solutions.
%{\bf Please check this description!}
%{\it An example for the case of $a=-0.68,\ C_0=0.9$ is illustrated.
%It has two real roots of $R$, the one is $R=-1.1235$ and the other
%$R=0.6535$. The dash-dotted lines are $v_c(x)$ related to the
%solutions. }
}\label{new1f}
\end{figure}
This solution gives the following scenario: the central region is
occupied by a high-density core with an outside medium of constant
density moving outward. Once the inner core begins to move towards
the centre, the outer part decelerates to 0 and also begins to move
inwards. We also see that the mass density decreases rapidly as $x$
becomes large. Figure \ref{new1f} shows a pair of such examples.

\subsection{Solutions crossing the Straight Sonic Critical Line Smoothly}

Only eigensolutions along the sonic critical line, derived in
subsection {\it 4.3.2}, can cross the sonic singular surface
smoothly. To investigate their properties, we start numerical
integration from the vicinity of a sonic critical point with initial
values given by one of the relevant eigensolutions. The analysis
here becomes much simpler in light of the scaling invariance
transformation (\ref{scaleinvariance}) and properties for every
sonic critical point are the same and hence an arbitrary point $x$
can be chosen for a certain pair of $a$ and $C_0$ parameters. By
solving for eigensolutions along the sonic critical line and then
extending the eigensolutions globally by numerical integrations, we
find that the two types of eigensolutions have qualitatively
different properties. A type I solution approaches the free-fall
asymptotic solution (\ref{fall}) in the inner part for small $x$ and
has an outer asymptotic solution described by solution
(\ref{desitterasymV}) and (\ref{desitterasymA}) at large $x$. The
physical scenario is that initially the gas with a constant density
has a tendency to move outwards, because of the thermal pressure
against gravity. The gravity force competes with the thermal
pressure and wins eventually as time goes on, and then the gas
begins to decelerate and accelerate to collapse towards the centre.
Now the self-gravity dominates the thermal pressure force completely
so that materials approach a free fall and finally smash onto the
central object.

\begin{figure}
\includegraphics[width=0.5\textwidth]{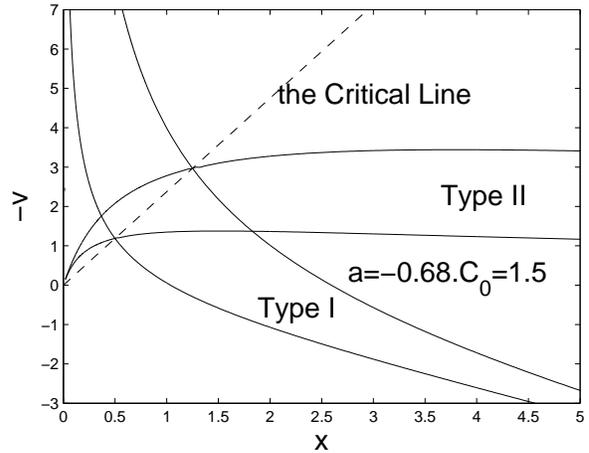}
\caption{
%{\bf comma after 0.68 in the figure.}
Two examples of semi-complete solutions crossing the straight sonic
critical line in two possible eigendirections with parameters
$a=-0.68$ and $C_0=1.5$. The sonic critical line is represented by
the straight dash-dotted line with a negative slope $k=-2.3754$. The
solid curves are the possible eigensolutions crossing the sonic
critical line, one of which is at $x=0.5$ and the other is at
$x=1.25$. The two eigensolutions approach different inner and outer
asymptotic solutions described previously. Because of the scaling
invariance of the ODEs, these two eigenproblems and the
corresponding eigensolutions are actually the same. Take the
eigenproblem at $x=1.25$ for example. Type I solution approaches
equation (\ref{fall}) with $m(0)=38.4$ near the centre and has an
outer asymptotic solution described by equation
(\ref{desitterasymA}) and (\ref{desitterasymV}) with parameters
$\beta=-1.6204$, $E=0.086$ and $F=-0.785$. Type II solution
approaches equation (\ref{new1}) at small $x$ with parameters
$\lambda=-3.00$, $R=-7.90$ and $N=0.179$. For large $x$, its
behaviour can be represented by equation (\ref{asym1}) with
$H=2.055$ and $L=-11.392$. Please note that for $a>-1$, $-v(x)$ of a
Type II solution always vanishes at large $x$; in this case of
$a=-0.68$, the leading term of $-v(x)$ at large $x$ scales as
$x^{-0.47}$.
%{\bf Please clarify!}
}\label{fig3}
\end{figure}

\begin{figure}
\includegraphics[width=0.5\textwidth]{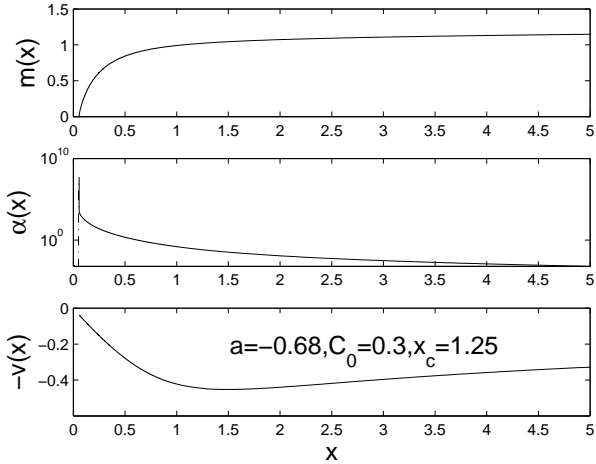}
\caption{%{\bf Vertical axis $m(x)$!} {\bf Use larger fonts.}
When $C_0$ is below $2^{4/3}/6\cong 0.42$, the asymptotic solution
(\ref{new1}) no longer exists. The type II eigensolution is
truncated before $x=0$. This figure shows an example of such
behaviour. The relevant parameters are $a=-0.68$ and $C_0=0.3$
with the corresponding slope $k=0.3571$ for the sonic critical
line. We obtain eigensolutions across the sonic critical line at
$x_c=1.25$. Apparently, the enclosed mass becomes 0 before $x$
reaches the origin, i.e., in the $-v$ versus $x$ diagram, the
velocity curve meets the line of $ax+v=0$ where the solution is
truncated. }\label{eigen2}
\end{figure}

In contrast, a type II solution has a quite different behaviour. In
the vicinity of the origin, the velocity vanishes while the mass
density diverges as described by asymptotic solution (\ref{new1})
around small $x$, while flow behaviour at large $x$ can be described
by asymptotic solution (\ref{asym1}).
%Based on the sign of slope $k$
%of the straight sonic critical line, type II solutions can be
%further subdivided into three subtypes. Subtype I: the slope $k$ is
%negative.\footnote{Note that the vertical axes in all figure
%presentations for radial velocities are $-v(x)$. Thus a negative
%value of slope $k$ corresponds to a straight critical line upward
%with increasing $x$.} Straightforward numerical integrations show
%that there is an inflow at infinity and under the gravity, the flow
%accelerates towards the centre. However, due to the increase of mass
%density, the thermal pressure increases gradually to exceed the
%gravity. The flow begins to decelerate and eventually the velocity
%vanishes at the centre. A negative slope $k$ corresponds to a larger
%constant $C_0$, leading to a larger thermal pressure against
%gravity. Subtype II corresponds to the marginal case of $k=0$, where
%type II solution is exactly the static SPS solution (\ref{static})
%and (\ref{staticcond}).
%{\bf Precise meaning?} {\bf Possible branch-off solutions
%with two $v_1$'s to a central free-fall solution?}
%As already discussed earlier, polytropic EWCS with $\gamma=4/3$ can
%be constructed here by connecting two branches of the two
%eigensolutions as shown in Fig. \ref{ewcs}. Subtype III is for a
%positive $k$. The flow has an outward velocity under the net force
%of thermal pressure and gravity. For a positive $k$, the asymptotic
%solution (\ref{new1}) does not exist.
Based on the value of $C_0$ compared with the critical value of
$2^{4/3}/6$, which determines whether $R$ has real roots, Type II
solutions can be divided into two subtypes. Subtype I: For $C_0\geq
2^{4/3}/6$ corresponding to the existence of real roots of $R$, Type
II solutions will approach $x=0$ as described by asymptotic solution
(\ref{new1}). A special case is the SPS solution with $\gamma=4/3$
when $k=0$. One can prove that the value of $C_0$ for $k=0$ is not
smaller than $2^{4/3}/6$. As discussed earlier, EWCS with
$\gamma=4/3$ can be constructed here by connecting two branches of
the two eigensolutions along the sonic critical line with $k=0$.
These subtype solutions describe the following scenario. The outer
part of the fluid system has a common flow behaviours which can be
an inflow, or an outflow, or even a static envelope, while in the
inner part, the pressure force and self-gravity compete with each
other such that the magnitude of the radial speed remains finite and
eventually vanishes at the centre. Subtype II: $C_0<2^{4/3}/6$ so
that asymptotic solution (\ref{new1}) does not exist. A numerical
integration backwards would be truncated before $x$ becomes
sufficiently small. Physically, the enclosed mass $m(x)$ is related
to the factor $ax+v$ by equation (\ref{meq}). Thus if $v(x)$ curve
occasionally approaches the straight line $ax+v=0$, then there is no
material inside this `radius' $x_v$. In other words, a spherical
void surrounds the centre and expands as time goes on in a
self-similar manner. For $a<-1$ cases, the boundary of such a
spherical void expands with deceleration and the edge radius is
proportional to $t^{-a-1}$. From ODEs (\ref{eq1})$-$(\ref{eq5}), the
behaviours of density, velocity and pressure can be deduced. The
enclosed mass within that point is zero, the reduced density
$\alpha$ is finite, the reduced pressure $p$ approaches zero there,
and the pressure gradient $dp/dx=-a(a+1)\alpha$ remains finite
according to equation (\ref{eq3}). Equation (\ref{eq4}) requires the
following limit
\begin{eqnarray}
\lim_{ax+v\rightarrow0}\frac{(ax+v)}{\alpha}
\frac{d\alpha}{dx}=-\frac{2(2+a-\gamma)}{\gamma}\ ,
\label{voidlimit}
\end{eqnarray}
and it follows from equation (\ref{eq5}) that
\begin{eqnarray}
\frac{dv}{dx}=2(1+a)+\frac{2(2+a-\gamma)}{\gamma}\
\end{eqnarray}
remains finite there.
%{\bf Please clarify!}
Numerical results also confirm the situation that the enclosed mass
$m(x)$ becomes $0$ before $x$ reaches the origin. An example of
$a=-0.68$ and $C_0=0.3$ is shown in Figure \ref{eigen2}
%{\bf Other parameters? In which figure?}
for the reduced quantities, such as $m(x)$, $\alpha(x)$ and $-v(x)$
in top, middle, and bottom panels respectively.
%of two different eigensolutions. {\bf `two', Meaning?}
%{\bf Please discuss solution properties around $m(x)=0$,
%e.g., other relevant variables.}
This shows the real possibility of a spherical void occupying the
central region of a certain astrophysical system (e.g., clouds,
bubbles, planetary nebulae, stars or supernova remnants etc.) during
its evolution under joint action of thermal pressure and
self-gravity. Previously, Goldreich \& Fillmore (1984b) discussed
collisionless particles with self-gravity in an Einstein-de Sitter
expanding universe. Steep perturbations can give rise to voids
surrounded by overdense shells with sharp edges. Our preliminary
results here show that in addition to the expansion of universe, a
spherical matter system with thermal pressure against self-gravity
can also lead to the formation of a central spherical void with an
overdense shell along a sharp edge.

\subsection{Self-Similar Flow Solutions with Shocks}

Global behaviours of eigensolutions crossing the sonic critical line
have been explored numerically. Starting from the two eigensolutions
on the sonic critical line and integrating towards small $x$, one
will approach the free-fall asymptotic solution (\ref{fall}) and the
other will approach the new asymptotic solution (\ref{new1}) (see
solution examples in Fig. \ref{fig3}). Type II solutions in Fig.
\ref{fig3} touch the sonic critical line twice. Other than this
special situation, due to the scaling invariance property, we are
unable to construct any global solutions across the sonic critical
line twice smoothly which are possible in the isothermal cases of
Lou \& Shen (2004) and the conventional polytropic cases of Lou \&
Wang (2006).
%{\bf Need a discussion on this!}

In this subsection, we turn our attention to self-similar flows with
shocks. From now on, subscripts $1$ and $2$ represent upstream and
downstream sides of a shock, respectively. In particular, we use
$C_1$ and $C_2$ to represent $C_0$ of the upstream and downstream
sides, respectively. Because it involves local sound speed with
respect to the shock reference framework in both upstream and
downstream sides, we also calculate the corresponding sound speed
%$c$ with %{\bf P or p here?}
$c_s\equiv (\gamma P/\rho)^{1/2}=(\gamma
C_0m^{2/3}\alpha^{1/3})^{1/2}$ for each branch of solutions.

%{\bf In reference to Figure 9, shock connection to SPS solution?}

We begin with free-fall core collapse solutions.
\begin{figure}
\includegraphics[width=0.5\textwidth]{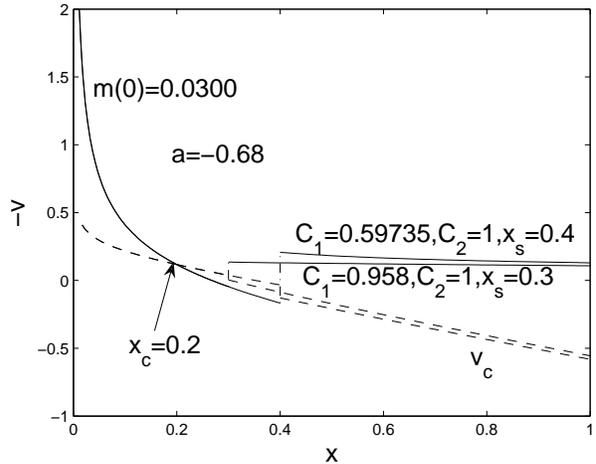}
\caption{%{\bf Use larger fonts!} {\bf More exploration needed}
Two shock flow solutions are illustrated here. These solutions
connect free-fall asymptotic solution at the inner part (small
$x$) with an inflow or contraction at infinity (large $x$). The
solid curves in the figure presents solutions and the dashed
curves are corresponding curves of $v_c$ defined in equation
(\ref{defvc}). The scaling index $a=-0.68$. The two solutions
share the same downstream branch with the free-fall solution
parameter $m(0)=0.030$ and $C_2=1.00$ and it crosses the sonic
critical line smoothly at $x_c=0.2$. The upstream side shows an
inflow described by equation (\ref{asym1}), with a set of
parameters $\{C_1,x_s,H,L\}$.
%{\bf You mean $C_1$?} confirming one branch.
Two example solutions correspond to
$\{0.5973,0.40,0.0019,-0.1322\}$ and
$\{0.9580,0.30,0.0019,-0.1246\}$.
%For one shock, all relevant
%parameters can be written in one set $\{a,\ C_1,\ C_2,\ m(0),\ x_0,\
%x_s,\ L,\ H\}$. {\bf Define notations!} The outer part is upstream
%while the inner part is downstream. The parameter set for solid
%curves marked with $1$ is $\{-0.68,\ 0.5972,\ 1.00,\ 0.03,\ 0.2,\
%0.4,\ 0.00194,\ -0.1885\}$ and that for the solid curves marked with
%$2$ is $\{-0.68,\ 0.9580,\ 1.00,\ 0.03,\ 0.2,\ 0.3,\ 0.00191,\
%-0.1338\}$. The dash-dotted curves represent the local sound speed
%$v_c$. {\bf Negative sound speed?} {\bf Consistency of notations and
%statements in the text?}
}\label{shock1}
\end{figure}
From the discussion of collapse solutions without crossing the sonic
singular surface, any of this kind solution will cross the sonic
singular surface even number of times, either smoothly or by shocks.
By inspecting this topological characteristics and considering the
simplest case of a single shock, we infer that this type of
solutions, with shock jumps across the sonic singular surface,
should be also possible to cross the sonic critical line smoothly at
some critical point.
%{\bf Please clarify!}
Based on this observation, we specify a type I eigensolution at a
given sonic critical point and integrate away from it in both
directions. Let us take solutions shown in Figure \ref{shock1} as
examples of illustration. In the comoving reference framework of the
shock, the outer part is supersonic and is thus the upstream side,
and the inner part is the downstream side. Here we apply the
matching procedure in the $\alpha-v$ phase diagram introduced by
Hunter (1977). A notable differece from the case of $\gamma\neq4/3$
is that the value of $C_0$ and the shock position will affect the
value of $C_0$ and thus the sonic singular surface in the other
solution branch. We also have a considerable freedom to construct a
shock in one solution at a chosen place and then integrate forward.
Such a numerical solution may approach a certain asymptotic
solution, or it may encounter the sonic singular surface.
%If the case is the former one, a shock solution is then found.
Remember that when a numerical integration is from the downstream
side to the upstream side, the square of the Mach number ${\cal
M}_2^2$ of the downstream side should be within the range of
$(0.125,1)$ and thus the shock position $x_s$ must be in a
corresponding range. The $x_s$ value in the example of Figure
\ref{shock1} is around $0.4578$. This kind of gravitational core
collapse solutions together with other solutions investigated
previously, such as Shu (1977) and Lou \& Shen (2004), may describe
a possible stage of star formation in molecular clouds.
%{\bf Please clarify the above discussion.}
%{\bf need a bit more discussion}

Tsai \& Hsu(1995), Shu et al. (2002) and Bian \& Lou(2005) connected
the outer singular isothermal sphere (SIS) solution with either LP
type solution or free-fall solution in the inner region by shocks in
an isothermal gas. Using the matching procedure in the $\alpha-v$
phase diagram, shock flow solution of this kind with the free-fall
asymptotic solution at small $x$ also exists in our polytropic case
of $\gamma=4/3$.
\begin{figure}
\includegraphics[width=.5\textwidth]{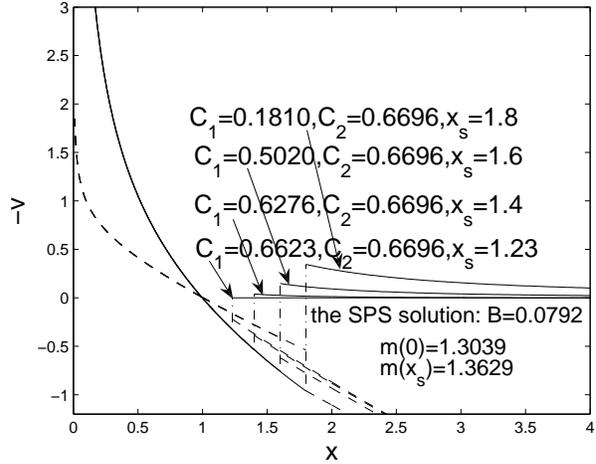}
\caption{Examples of inner free fall and shock jumps to the SPS
outer part and other asymptotic flow solutions far away. For the
SPS shock connection, the shock is located at $x_s=1.23$ with
$C_1=0.6623$ and $C_2=0.6696$ and the inner solution crosses the
sonic singular surface again at $x_c=0.992$ smoothly. The mass at
the centre is $m(0)=1.3039$ and the mass enclosed within the shock
front is $m(x_s)=1.3629$. For the three outer asymptotic flow
solutions from the top in order, we have relevant shock parameters
$\{C_1,\ C_2,\ x_s\}$ to be $\{0.1810,\ 0.6690,\ 1.8 \}$,
$\{0.5020,\ 0.6690,\ 1.6\}$, $\{0.6276,\ 0.6690,\ 1.4\}$,
respectively, and relevant flow parameters $\{H,\ L,\ K\}$ of
asymptotic solution (\ref{asym1}) to be $\{0.082,\ -0.1,\ -1.0\}$,
$\{ 0.080,\ 0.0,\ -0.3\}$, $\{0.080,\ 0.0,\ -0.07\}$, respectively
(see asymptotic solution \ref{asym1}).
%
%Other relevant parameters are also shown in the figure.
%{\bf shocks with breezes or outflows?}
%{\bf Parameters at large $x$?}
}
\end{figure}
This particular kind of shock solutions depicts the following
scenario. Initially the outside gas is in a radial force balance and
the collapse starts from the central core region. Effects such as
changes in the centre propagates outwards in the form of a
self-similar shock. Materials are blown out by this shock. Because
the gravity is stronger than the pressure force, materials
eventually stop moving outwards and fall towards the centre.

\begin{figure}
\includegraphics[width=0.5\textwidth]{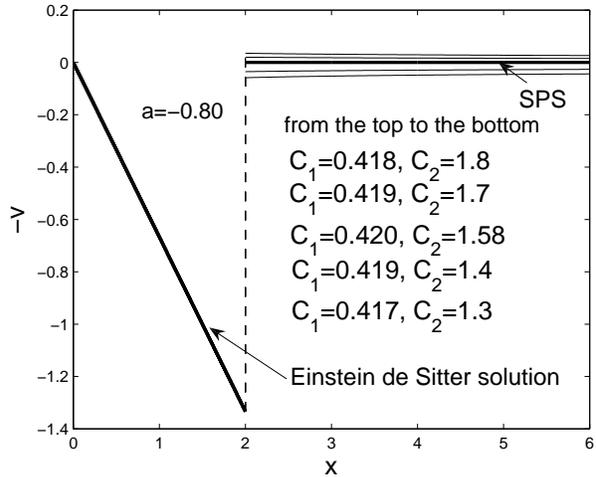}
\caption{
%{\bf Caption for Einstein-de Sitter with other asymptotic soltuions.}
This figure illustrates a special case (the heavy solid line) of
shock solution which connects the inner Einstein-de Sitter
solution and the outer SPS with $\gamma=4/3$. Here, $C_2=1.58$ and
the corresponding value of $a$ is $-0.80$. This kind of solutions
may be regarded as a limiting solution as $C_2$ approaches $1.58$
while keeping $a$ the same. Also shown in this Figure are
different upstream solutions of $C_2=1.3,\ 1.4,\ 1.7,\ 1.8$,
respectively. For these four upstream solutions away from the SPS,
the three relevant parameters $\{H,\ K,\ L\}$ for asymptotic
solution (\ref{asym1}) are $\{0.134,\ -0.007,\ 0.068\}$,
$\{0.126,\ 0.00,\ 0.041\}$, $\{0.106,\ 0.00,\ -0.024\}$ and
$\{0.101,\ 0.00,\ -0.041\}$, respectively.}\label{EdSshock}
\end{figure}
While all LP type asymptotic solutions degenerate to the Einstein-de
Sitter solution in our case of $\gamma=4/3$, shock solutions can
also be constructed to connect the inner Einstein de Sitter solution
with an outer SPS (see Fig. \ref{EdSshock}). Naturally, the outer
SPS part is the upstream side with $v_1=0$, and the inner
Einstein-de Sitter solution is the downstream side with $v_2=2x_s/3$
and $\alpha_2=2(1+2\sqrt[3]{3}C_2)^{-1}/3$ at $x_s$ where $C_2$ is
set to an appropriate value. Using $v_2$, $\alpha_2$ and $C_2$, we
can express $v_1$ in terms of the scaling index $a$. The condition
of $v_1=0$ then appears as a quadratic equation
%{\bf the explicit form?}
of $a$, which needs to be solved for $a$ with $a<-2/3$. Once
this is done, we use $v_2$, $\alpha_2$, $C_2$ and the relevant
root(s) of $a$ to calculate the Mach number ${\cal M}_2$ on the
downstream side to check whether the requirement $1/8<{\cal M}_2<1$
is met. Once everything is complete and consistent, one of this kind
of shock solutions is then constructed. We show an example here in
Figure \ref{EdSshock} with $C_2=1.5,\ a=-0.799,\ \alpha_2=0.1252$,
leading to $C_1=0.383,\ \alpha_1=0.0207$ correspondingly, where the
self-similar shock position $x_s$ can be any positive values because
of the scale invariance property.
%{\bf It should be possible to also connect inner Einstein-de
%Sitter solution with shocks to outer flow solutions? Please
%show a few examples.} {\bf Need parameters.}

Shocks can also be inserted to connect the only two analytic
solutions available, namely, the static SPS solution outside and
the Einstein-de Sitter solutions inside. To construct this kind of
shock solutions, the upstream quantities are $v_1=0$ and
$\alpha_1=Bx_s^{2/a}$ with $C_1$ (i.e., the upstream $C_0$)
satisfying conditions (\ref{static}) and (\ref{staticcond}) for
the existence of SPS,
%(see equation \ref{staticcond})
while the downstream quantities are $v_2=2x_s/3$ and
$\alpha_2=2/[3(1+2\sqrt[3]{3}C_2)]$ from the Einstein-de Sitter
solution (\ref{eqq}). With these constraints, it is
straightforward to determine $C_2=1.58$,
%{\bf You mean 1.58 not 0.58?}
$C_1=0.42$ and $a=-0.80$. Figure \ref{EdSshock} shows this
solution with the shock position at $x_s=2.0$. It is easy to see
from the figure that this solution represents the limiting
solution of a solution family of $C_2$ approaching $1.58$ with
$K=0$. It is unlike the isothermal results of Tsai \& Hsu (1995)
where this kind of solutions is a limit of a family of breeze
solutions (Shu et al. 2002). Instead, it is a critical state to
distinguish asymtotic outflow and inflow solutions. For $C_2$
being slightly larger than $1.58$, the asymptotic solution
represents an inflow, while for $C_2<1.58$ the asymptotic solution
corresponds to an outflow. In fact, this Einstein-de Sitter shock
model can be applied to an explosion process with a stellar
interior as an alternative of the rebound shock model of Lou \&
Wang (2006, 2007) described at the beginning of the next
paragraph. The major difference here is a constant density within
the shock front instead of being a diverging density near the
centre; outside the shock front, the density approaches a
power-law scaling with either infalling or outgoing stellar
materials. When this shock front reaches the photosphere of the
progenitor, we start to see observable effects of a supernova in
optical bands.

Lou \& Wang (2006) utilized a self-similar polytropic model to
construct the gravitational core collapse and rebound shock
processes in supernova explosions. Their conventional polytropic
model solution with $\gamma\neq 4/3$ is to connect quasi-static
solutions at small $x$ with outer asymptotic flow solutions at
large $x$ by outgoing shocks. That model was recently generalized
to include a random magnetic field using a magnetohydrodynamic
(MHD) approach and to explore the origin of strong magnetic fields
of compact objects (Lou \& Wang 2007; Wang \& Lou 2007). Based on
our model framework here, this can also be done for a general
polytropic gas with $\gamma=4/3$ and thus $q=2/3$. Starting
numerical integrations both from the centre and from infinity
(actually a sufficiently large $x$) and choosing a proper meeting
point to match solutions in the $\alpha-v$ phase diagram, we
adjust the shock position and parameters of outer asymptotic
solution (\ref{asym1}) to construct sensible solutions (e.g., Lou
\& Shen 2004).
\begin{figure}
\includegraphics[width=0.5\textwidth]{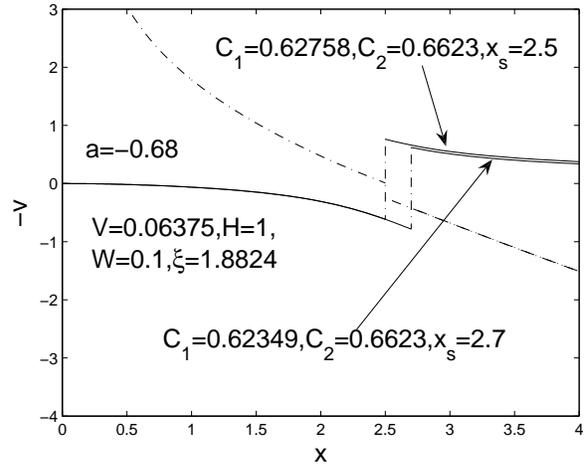}
\caption{%{\bf Please use larger fonts.}
Shock solutions with quasi-static asymptote of equation
(\ref{quasi1}) and (\ref{quasi2}) at small $x$ are shown here. The
power index parameter $a$ is $-0.68$. In the downstream side, we use
the same quasi-static solution while two uptream branches are
different. The parameters in the downstream side for the
quasi-static solution are $C_{2}=0.6623,\ B=1.00,\ W=0.10,\
V=0.06375,\ \xi=1.8824$.
%{\bf Change $H$ to $B$ in the figure.}
The upstream branch converges to asymptotic solution (\ref{asym1})
at large $x$. Thus each upstream solution can be identified by a set
of parameters $\{C_1,\ x_s,\ H,\ L\}$, where $x_s$ is the shock
location and the last two are the coefficients in asymptotic
solution (\ref{asym1}). In this illustration the two sets of
parameters are $\{0.62758,\ 2.5,\ 1.0096,\ -0.5323\}$ and
$\{0.62349,\ 2.7,\ 1.0149,\ -0.4486\}$. The dash-dotted curves
%{\bf Please check the linetype.}
are corresponding segments of $v_c$
curve defined by condition (\ref{defvc}).}
\end{figure}
Alternatively, instead of the above matching procedure, we can
also start a numerical integration from the vicinity of the centre
and then choose a certain point as the shock location. Shock jump
conditions (\ref{cond1})$-$(\ref{cond3}) determine all physical
variables on the upstream side of a shock. A further numerical
integration outwards until $x$ is sufficiently large completes the
solution construction procedure. Using the numerical solution thus
obtained, we can match with asymptotic solutions to determine
relevant parameters. We find that the self-similar shock position
can only exist within a finite interval of $x$ (e.g., in the case
of $a=-0.68$, the self-similar shock position falls within the
range of $1.3\leq x_s\leq4.23$). Outside this interval of $x$, the
square of Mach number ${\cal M}_2^2$ on the downstream side would
be smaller than $1/8$ which is unphysical by our analysis on
self-similar shock conditions. This $\gamma=4/3$ rebound shock
model for supernovae may be more appropriate in certain aspects.
During the initial phase for the emergence of a rebound shock in
the dense stellar core, neutrino pressure, radiation pressure and
gas pressure together may be modelled by a polytropic mixture of
$\gamma=4/3$. The diverging density near the centre is expected to
create a highly degenerate core there.

We also construct shocks connecting asymptotic solutions
(\ref{asym1}) at infinity and (\ref{new1}) near the centre.
\begin{figure}
\includegraphics[width=0.5\textwidth]{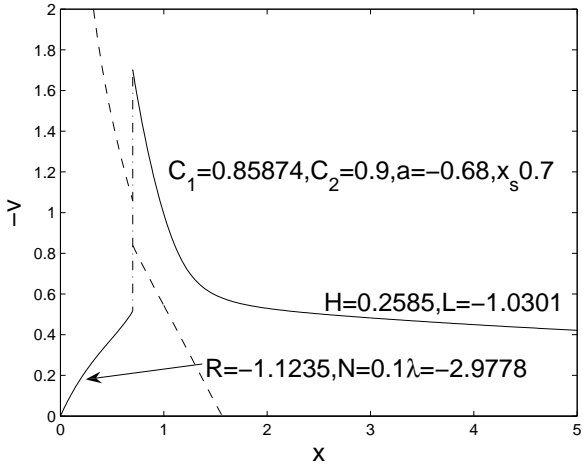}
\caption{A similarity shock solution with $a=-0.68$ and
$C_{1}=0.85874,\ C_{2}=0.9$
%{\bf Notation definitions! Somewhat cumbersome. }
is illustrated. It connects asymptotic solution (\ref{asym1}) with
parameters $H=0.2585,\ L=-1.0301$ and asymptotic solution
(\ref{new1}) with parameters $R=-1.1235,\ N=0.1,\
\lambda=-2.9778$.
%{\bf Change $M$ to $R$ in the figure.}
%{\bf Notation definitions!}
The solid curve represents the shock solution and the dashed curve
represents corresponding segments of $v_c$ curve defined by
condition (\ref{defvc}).
%{\bf Notation definition!}
%which also expresses the relative position of the solution and the
%singular surface. {\bf Please clarify!} {\bf Need a bit more
%exploration.}
}
\end{figure}
In the above analysis, we know that the inner part of this solution
can only appear in the first quadrant in the plane of $-v$ versus
$x$. The numerical treatment starts from the centre and goes
outwards. Before the solution meets the sonic singular surface, jump
conditions are included to introduce a shock across the sonic
singular surface. We then continue to integrate outwards until $x$
is sufficiently large to match with asymptotic solutions at large
$x$. Numerical experiments show that almost all such solutions match
asymptotic solution (\ref{asym1}), in which mass density and flow
velocity both converge. It is also possible that after a shock
jumping across the sonic singular surface and integrating outwards,
the radial flow velocity decreases rapidly so that it crashes onto
the sonic singular surface again.
%In general, it is not an eigensolution for any critical point
%and thus this solution is aborted. {\bf Precise meaning?}
In our numerical experiment, we do not find solutions with twin
shocks or others, which jump across the singular surface twice or
more (see Bian \& Lou 2005).

\section{Discussion and Conclusions}

We have explored and examined the similarity flow solution
structures in a general polytropic gas with a polytropic index
$\gamma=4/3$. Previously, Goldreich \& Weber (1980) considered a
special case of $\gamma=4/3$ with a constant specific entropy. By
their assumptions and analysis, only homologous collapse solutions
exist by invoking the time reversal invariance, i.e., the radial
flow velocity $u$ takes on the form of $2r/(3t)$ until the mass
density vanishes at a certain point. Yahil (1983) mentioned the case
of Goldreich \& Weber (1980) as a special limit. In reference to
earlier work and based on a self-similar transformation, we
systematically examined the case of $\gamma=4/3$ with specific
entropy conservation along streamlines. We have substantially
generalized the earlier analyses, discovered new asymptotic
solutions, and constructed various self-similar solutions without or
with shocks.

In reference to earlier analyses of Goldreich \& Weber (1980) and
Yahil (1983), our model framework mainly focuses on $\gamma=4/3$
with the conservation of specific entropy along streamlines, which
is more general and perhaps, closer to reality than the conventional
polytropic gas of a constant specific entropy everywhere at all
times. Of course, the case of a constant specific entropy is also
possible and can be properly accommodated and treated within our
polytropic model framework of $\gamma=4/3$. Under our more general
formalism, we extend the work of Goldreich \& Weber (1980) and
obtain many interesting results. The solutions are divided into two
broad classes: solutions with $a=-2/3$ precisely belong to Class I
and solutions with $a<-2/3$ belong to Class II. For the situation of
$-2/3<a<0$ as mentioned at the beginning of deducing asymptotic
behaviours, the divergent velocity at large $x$ is not of interest
%is not suitable for an infinite solution for our concerning problem
and hence we only consider two classes I and II solutions.
%{\bf You classification notations may be confusing with your other notations!}
%{\bf How about the $-2/3<a<0$ case?}

Class I solutions are characterized by $P\propto\rho^{4/3}$ with the
proportional coefficient related to the specific entropy being an
arbitrary function of $x$, while for Class II solutions, this
proportional coefficient depends on the enclosed mass $M$ in a
power-law form. We discuss these two classes separately.

\subsection{Class I Self-Similar Solutions}

Class I self-similar solutions represent a substantial extension of
the special solutions with a constant entropy derived by Goldreich
\& Weber (1980). For an astrophysical system such as stars, the
specific entropy is not expected to be a global constant in general.
For a stellar interior, this depends on the competition between
thermal kinetic energy and Fermi energy as determined by the mass
density. Qualitatively speaking, especially for a compact object,
the closer to the centre, the closer the material is in a degenerate
state; this would correspond to a smaller specific entropy. However,
the density is relatively small and the temperature is relatively
low in the outer part of a star, perhaps also leading to a lower
level of entropy. We do not yet know the exact distribution of
specific entropy within a star so far. Thus the case of a constant
entropy is the simplest to consider and provides a certain sense for
a homologous dynamic process. The model analysis of this paper is
more general and allows for a fairly arbitrary distribution of
specific entropy along streamlines. Meanwhile, the radial velocity
profile remains always equal to $2r/(3t)$. For a given time $t$, the
radial velocity increases linearly with increasing radius $r$.
Hence, this solution can be valid within a finite radial extent. It
turns out that the mass density vanishes at some place referred as
the outer boundary of the flow system.

According to the model analysis of Goldreich \& Weber (1980), a
pre-collapse progenitor star of a static configuration may evolve
into a homologous core collapsing phase (see Figure \ref{GWf}), when
the pressure suddenly decreases by a fraction within a range of
$\sim 2.9\%$.
%, the system will enter the homologous collapsing phase, which,
Early simulations of Bethe et al. (1979) indicated a substantially
larger pressure reduction of $26\%$ is needed in order to initiate
collapse in supernova explosions. The much smaller fraction change
of pressure reduction for a homologous core collapse given by
Goldreich \& Weber (1980) is actually related to the assumption of a
constant specific entropy (i.e., their constant $\kappa$) in space
and time. Requiring specific entropy conservation along streamlines
and allowing the specific entropy to be a function of space and
time, it is possible to have a homologous core collapse for a much
larger fractional change of pressure reduction.
%They explained this inconsistency by introducing a less massive
%core which would experience homologously collapse. But
In our more general analysis and notations, we find that other forms
of $g(x)$ instead of $g(x)=1$ can give rise to a fractional change
of $26\%$ or larger for a pressure reduction. Physically, this
corresponds to different distributions of specific entropy along
streamlines.

To illustrate this case specifically, we choose $g(x)=1/(1+\epsilon
x)$ where $\epsilon>0$ is an adjustable parameter to gradually
modify the shape of $g(x)$. When $x$ is sufficiently small, $g(x)$
is nearly equal to $1$, analogous to the $g(x)=1$ case. Globally
$g(x)$ is a decreasing function with increasing $x$. When $\epsilon$
is small, $g(x)$ decreases slowly and only deviates from $g(x)=1$
case when $x$ is sufficiently large. For large $\epsilon>0$, the
result will differ considerably from that of Goldreich \& Weber
(1980). We carry out such a $g(x)$ experiment numerically.

Substituting dimensional quantities into the dimensionless state
function $p=g(x)\alpha^{4/3}$,
%{\bf I replace $\rho$ by $\alpha$ here.}
the dimensional equation of state can be written explicitly as
\begin{eqnarray}
P=\frac{g(x)}{A^2}(4\pi G)^{1/3}\rho^{4/3}\ ,
\end{eqnarray}
where for a given time $t$, $g(x)$ corresponds to a radial
distribution of specific entropy. Parameter $A$ also varies for
different values of chosen $f(0)$ for a certain system in which the
total enclosed mass $M$ is conserved and thus the value of $A$
parameter can be deduced from equation (\ref{GWM}). The variation in
$A$ value actually corresponds to the variation range for fractional
change in pressure denoted by $r_p$; in Goldreich \& Weber (1980),
this $r_p$ is $2.9\%$ as $f(0)$ increases from $f_c$ to infinity.
For any given form of $g(x)$,
%as GW show in their paper,
the limiting case is the Lane-Emden equation (e.g., Chandrasekhar
1939) as long as $g(x)\rightarrow1$ as $x\rightarrow 0^+$. Of
course, this applies to our chosen form of
%So is the limit case of
$g(x)=1/(1+\epsilon x)$ as $x\rightarrow 0^+$.

\begin{table}
\begin{center}
\caption{When $g(x)$ takes the form of $1/(1+\epsilon x)$, the
results are shown in Figure \ref{GWexf}. As $f(0)$ is sufficiently
large, the outer boundary at a finite $x$ where $f(x)=0$ is
extremely small and thus $g(x)$ can be almost treated as a
constant; these results vary little as compared with those of
Goldreich \& Weber (1980) for $f(0)\rightarrow+\infty$
corresponding to the Lane-Emden equation. Results in Figure
\ref{GWexf} mainly focus on the limiting case of $f(0)\rightarrow
f_c^+$.
%{\bf You mean the minimum of $f_c^+$? Please clarify your statements.}
For possible $f(x)$ solutions of a homologous core collapse, we
use $r_p$ to denote the range for fractional change of pressure
variation as $f(0)$ increases from $f_c$ to infinity.
%{\bf Please define notation $r_p$.}
}\label{GWext}
\begin{tabular}{c|c|c|c}
 $\epsilon$ & $f_c$ & $\bar{\rho}/\rho_c$ at $f_c$ &
 $r_p$ \\
 \hline
 $0$ & $4.67047$ & $0.00655$ & $2.9\%$ \\
 $0.01$ & $4.58642$ & $0.00693$ & $3.6\%$ \\
 $0.05$ & $4.28755$ & $0.00848$ & $6.4\%$ \\
 $0.1$ & $3.98354$ & $0.0106$ & $10\%$ \\
 $0.3$ & $3.23130$ & $0.0198$ & $26\%$ \\
 $0.5$ & $2.83696$ & $0.0293$ & $42\%$ \\
 \hline
\end{tabular}
\end{center}
\end{table}

\begin{figure}
\includegraphics[width=0.5\textwidth]{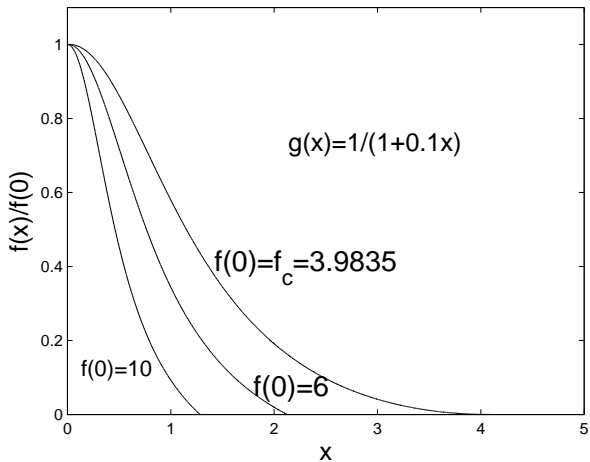}
\caption{Results of the normalized $f(x)/f(0)$ is displayed for
$g(x)=1/(1+0.1x)$ and $\epsilon=0.1$. In this case, the minimum of
$f(0)$ is $f_c=3.9835$ for physical solutions. }\label{GWexf}
\end{figure}

Table \ref{GWext} and Figure \ref{GWexf} show major results for a
range of $\epsilon>0$ values. Numerical experiments indicate that as
$\epsilon$ increases from $0$, the limiting value $f_c$ of $f(0)$
decreases while the density ratio $\bar\rho/\rho_c$ increases. More
importantly, the range of fractional change $r_p$ by which the
pressure can be reduced for a homologous core collapse becomes
larger and larger. Table \ref{GWext} shows that for $\epsilon=0.3$,
we have $r_p\cong 26\%$; in other words, for such a large fractional
change in pressure reduction (e.g., Bethe et al. 1979) in order to
initiate supernovae, it is still possible for a homologous core
collapse prior to the development of a rebound shock. It is
conceptually important that a different specific entropy
distribution of $g(x)$ from a constant value can lead to a better
agreement with numerical simulations; this appears more effective
than the inclusion of a less massive core in the centre as mentioned
in Goldreich \& Weber(1980).

\subsection{Class II Self-Similar Solutions}

In addition to extensions of Goldreich \& Weber (1980) discussed in
the above subsection, we also substantially generalize the
self-similar solution space for $\gamma=4/3$
%in a complementary manner
by adjusting the scaling index $a$. In contrast to $\gamma\neq 4/3$,
a straightforward analysis with $\gamma=4/3$ leads to an exact value
of $q=2/3$ that is independent of $a$. The dimensional equation of
state then takes the form of $P\propto M^{2/3}\rho^{4/3}$ with a
constant proportional coefficient. We can compare the thermal energy
$k_BT$, where $k_B$ is the Boltzmann constant, with the Fermi energy
$\varepsilon_F$. Neglecting the rest mass in the relativistic
regime, the relationship between total energy $\varepsilon$ and
momentum $\mathbbm{p}$ for a single particle can be written as
$\varepsilon=c\mathbbm{p}$ where $c$ is the speed of light, leading
to $\varepsilon_F\propto\rho^{1/3}$. In our model, the state
function gives $k_BT\propto P/\rho\propto M^{2/3}\rho^{1/3}$. It
follows that
\begin{eqnarray}
\frac{k_BT}{\varepsilon_F}\propto M^{2/3}\ .
\end{eqnarray}
%{\bf Typos around here?}
The enclosed mass $M$ is a non-decreasing function in radius $r$. At
a given time $t$, we see from this relation that, at small $r$, the
enclosed mass is small and hence this ratio is also small.
Physically in the inner core of a star, where materials are highly
condensed and may be close to a degenerate state, the specific
entropy is low.
%{\bf What is the point? Please explain. }

A modified self-similar transformation is introduced for
$\gamma=4/3$. In the self-similar transformation for a conventional
polytropic gas (i.e., $P=\kappa\rho^{\gamma}$ with globally constant
$\kappa$ at all times), the sound speed appears either explicitly or
implicitly. In contrast, we here use an integration constant $C_0$
relating to the sound speed and transformation (\ref{eq07}) does not
involve the sound speed because of the uniqueness of the
$\gamma=4/3$ case; this $C_0$ coefficient is in fact allowed by the
transformation and is an adjustable parameter in our analysis for
astrophysical applications. At a deeper level, we realize that as
the special self-similar transformation (\ref{eq07}) for
$\gamma=4/3$ does not involve the sound speed, we have a scaling
invariance (\ref{scaleinvariance}) which simplifies our theoretical
analysis considerably.

By comparisons and analogies of solutions known for $\gamma\neq4/3$,
we try our best to find the counterpart solutions and to discover
new solutions for $\gamma=4/3$. Global analytic solutions, i.e.,
static SPS solution (\ref{static}) and (\ref{staticcond}) and
Einstein-de Sitter expansion solution (\ref{eqq}), still exist for
$\gamma=4/3$ with some modifications. Analytic asymptotic solutions
of various kinds are also derived for both large and small $x$.
However, the LP-type solution no longer exists for $\gamma=4/3$
(thus $q=2/3$) except for rare situations, while other counterpart
solutions are readily found. In addition, a new type asymptotic
solution (\ref{new1}) is discovered in the regime of small $x$. It
seems that this type of asymptotic polytropic solutions only exists
for $\gamma=4/3$.

We have also examined properties of the sonic singular surface and
the sonic critical line of coupled nonlinear ODEs (\ref{eq11}).
%and (\ref{eq12}) and also the sonic critical line on the sonic singular surface.
A salient feature of $\gamma=4/3$ case is that all sonic critical
lines are straight and pass through the origin $x=0$ and $v=0$ in
the $-v(x)$ versus $x$ presentation; while first revealed by
extensive numerical experiments, these remarkable results can be
proven analytically. It is also fairly straightforward to derive two
eigensolutions to smoothly cross the straight sonic critical line.
In later analyses, we realize that the sonic critical line cannot
fully represents the behaviours of the sonic singular surface,
especially for constructing shocks. We hence use $v_c(x)$ defined by
equation (\ref{defvc}) for each solution, which is another curve on
the sonic singular surface and tightly relates to the current
solution, to show the interrelation between the current solution and
the sonic singular surface. According to definition (\ref{defvc}),
the solution meets the singular surface if and only if the solution
and the corresponding $v_c$ intersects; meanwhile, a shock solution
jumping across the sonic singular surface also jumps across the
corresponding $v_c$ in $v(x)$ versus $x$ plane. %{\bf Please clarify!}
The standard Runge-Kutta scheme (e.g., Press et al. 1986) is used to
numerically integrate coupled nonlinear ODEs (\ref{eq11}) and
(\ref{eq12}) to connect various asymptotic solutions and
eigensolutions across the sonic critical line. We also construct
possible solutions with and without shocks for potential
astrophysical applications. From the behaviours of these
semi-complete solutions, we can see that many such solutions have
similar behaviours as those for $\gamma\neq 4/3$ in a qualitative
manner. And our solutions can be sensibly regarded as limits of
$\gamma\rightarrow4/3$ when a polytropic gas becomes
relativistically hot or degenerate. Our analysis for $\gamma=4/3$
does share a certain common characteristics with cases for
$\gamma\neq4/3$.

%{\bf Please examine and discuss properties of
%void boundary carefully and systematically!}
We can readily solve for eigensolutions crossing the straight sonic
critical line. Once the slope $k$ of the sonic critical line is
positive, we can construct a new type of self-similar solution
characterized by an expanding central spherical void within which
the enclosed mass is zero or negligible. Around the edge of such an
expanding void, there exists an overdensed shell where the density
variation becomes rather steep. Diffusion processes are expected to
smooth out such relatively steep gradients locally. To consider
properties of spherical void boundary, we first note that such a
void expands with a radial speed
\begin{eqnarray}
 \dot{r}_e=-ar_e/t\qquad\quad\Rightarrow \qquad\quad r_e\propto
 t^{-a}\ ,
\end{eqnarray}
where $r_e$ stands for the radius of the spherical void edge. The
spherical void edge evolves as a power law of time $t$ with a
scaling index $-a$. One also notes by equation (\ref{voidlimit})
that the density gradient approaches a negative infinity near the
void edge. We expect that in a narrow region near the spherical void
edge, materials are actually diffused instead of being so sharply
distributed as shown by our solution mathematically. Within this
narrow region, the local evolution does not behave self-similarly
and may not be spherically symmetric, while the overall self-similar
profile remains on large scales. In the outer part, the mass density
scales as $\rho\propto r^{2/a}$ and the radial flow velocity remains
finite with a wind. In fact, it is also possible to construct
various shock solutions with a central void.

At this stage, we may outline a physical scenario in the context of
a supernova explosion. During the core collapse of a progenitor,
neutrons are formed in abundance and neutrinos of relativistic
energies are released. In a high-density environment, neutrino
opacity is extremely high so that neutrino pressure, radiation
pressure and gas pressure work together to drive the central core
expansion. In the relativistic regime, we may ignore tiny neutrino
masses and regard the neutrino gas as polytropic with an index
$\gamma_{n}=4/3$. Similarly, the radiation pressure resulting from
the photon gas trapped in the stellar interior can also be regarded
as polytropic with an index $\gamma_{\nu}=4/3$. In the hot stellar
core of high temperatures, we may approximate the thermal gas
pressure as polytropic with an index $\gamma_p\cong 4/3$. It might
be conceivable that under certain situations, the neutrino pressure
is so overwhelming such that a central void may start to form. As
the outer part expands and density drops, neutrinos escape while the
radiation and thermal gas pressures continue to drive the expansion.
It should be emphasized that in real situations, a grossly spherical
void may still encompass materials here and there but the mass
density inside is substantially lower than that of surroundings.

%How does this structure form?
%Another possible situation is, at some time, a perturbation occurs
%in a steady gas flow and creates a discontinuity around the centre
%within which little materials are left with particles of high energies.
%, such as neutrinos.
%All particles lying there before all flow outwards
%under the gradient of pressure. {\bf Meaning?}
%At the very beginning, the material outside is highly condensed and
%opaque to enclose the inner particles. In a much shorter time scale
%rather than the evolution time scale of the system, the material is
%cooled down and the optical depth decreases rapidly, in which
%duration, the inner particles with high energy escapes out, and thus
%the void forms.
In this context, we note the model work of Fillmore \& Goldreich
(1984b) who considered a collection of collisionless particles in an
expanding universe of the Einstein-de Sitter form. There is no
pressure effect from particles in their model. In essence, the
background Einstein-de Sitter expansion prescribed is similar to the
rapid expansion driven by the thermal pressure force in our model,
both providing the tendency for particles to move outwards in
competition with the inward self-gravity. The key physical
difference is that the Einstein-de Sitter expansion of the universe
is homogeneous (presumably driven by the ubiquitous dark energy)
while our gas expansion is driven by the thermal gas pressure
closely related to gas mass density and temperature. Not only in the
case $\gamma=4/3$, self-similar void solutions can also be
constructed for $\gamma\neq4/3$ and isothermal gas which we shall
investigate more thoroughly in separate papers.

Besides certain similarities with previous polytropic model analysis
with $\gamma\neq 4/3$, the case of $\gamma=4/3$ carries its own
unique features. First, because of scaling invariance
(\ref{scaleinvariance}), various self-similar solutions can be
readily classified, especially for the two eigensolutions across the
sonic critical line. Once solution properties at a chosen point $x$
have been examined completely, other points will have the same
solution characteristics by scaling invariance
(\ref{scaleinvariance}). This simplifies the analysis to a
considerable extent. Fundamentally, the cause of this scale
invariance (\ref{scaleinvariance}) is due to the fact that the sound
speed is not involved in self-similar transformation (\ref{eq07}).
In various solutions, the case of $\gamma=4/3$ also shows some
differences:
%EWCS {\bf Please check the $k=0$ situation!}
(i) the LP type solution does not exist, except for rare situations
(see Appendix \ref{LP}); (ii) when discussing the quasi-static
solution at small $x$, two sensible roots may be found for
$\gamma<4/3$ (see Lou \& Wang 2006), while one of the two roots is
always unacceptable for $\gamma=4/3$ with only one sensible root
being available in our model calculations; (iii) it is no longer
possible for a quasi-static solution at small $x$ to show a
vibration behaviour here. The solution quickly converges to an outer
asymptotic solution; and (iv) the sonic critical lines with constant
density are straight lines emanating from the origin in the $-v(x)$
versus $x$ presentation.

\subsection{Conclusions}

To sum up this paper, we have explored possible self-similar
solutions, both analytical and numerical, for a generalized
polytropic gas with $\gamma=4/3$. The classical analysis of
Goldreich \& Weber (1980) for a constant specific entropy everywhere
at all times is substantially extended by specific entropy
conservation along streamlines with specific entropy dependent on
time and space. This differs from what Yahil (1983) did in this
context.
%We invoke the conservation of specific entropy along streamlines and
%used the self-similar transformation of Fatuzzo et al. (2004).
In addition to counterparts of various previously known types of
polytropic solutions with $\gamma\neq 4/3$, we find two new
asymptotic solutions. One notable feature is that sonic critical
lines are all straight lines emanating from the origin in the
$-v(x)$ versus $x$ presentation with constant densities. Using all
asymptotic solutions available, two eigensolutions across the sonic
critical line and self-similar shock conditions, global
semi-complete solutions are constructed numerically.
%and tried to imagine corresponding physical process.
Two classes of self-similar solutions are investigated separately
according to the value of the scaling index $a$. For Class I
solutions with $a=-2/3$ precisely, we can simulate the homologous
evolution of a flow system once the distribution of specific entropy
is prescribed. These more general solutions for a homologous core
collapse (Goldreich \& Weber 1980) may be utilized to model the
dynamic formation of an inner compact core from a pre-collapse
stellar interior. Collapsing solutions of Class II with $a<-2/3$ may
also explain the formation of compact objects and other similar flow
systems, while expansion solutions with shocks can be utilized to
model supernova explosions (e.g., Lou \& Wang 2006, 2007).

By equations (\ref{eq1})$-$(\ref{eq5}), we emphasize several aspects
of solutions for $(3a+2)=0$ and for $(3a+2)\rightarrow 0$. Based on
our analysis, the Einstein-de Sitter solution with $\gamma=4/3$
exists for $(3a+2)\neq 0$, $(3a+2)\rightarrow 0$ and $(3a+2)=0$.
Except for this special Einstein-de Sitter solution, Class I
solutions valid for $(3a+2)=0$ cannot be obtained by taking the
limit of $(3a+2)\rightarrow 0$ for Class II solutions. In other
words, Class I and II solutions are qualitatively different
solutions and we need to consider them separately. By equations
(\ref{eq1}) and (\ref{meq}) during the limiting process of
$(3a+2)\rightarrow 0$, we must require $(ax+v)\rightarrow 0$ in
order to have a finite reduced mass $m(x)$. Only the Einstein-de
Sitter solution and Class I solutions with $(3a+2)=0$ bear this
unique feature for the reduced flow speed $v(x)$ while all other
Class II solutions are excluded by this limiting procedure.

In the course of investigation, we realize the possibility of
constructing self-similar solutions for dynamic evolution of central
spherical void in a flow system involving self-gravity and thermal
pressure. Here, the thermal pressure force drives the gas expansion
sufficiently fast and creates a central spherical void by
%provides similar effects like the expansion of the universe,
pushing materials outwards. By specific examples, we now prove by
analytical and numerical calculations that a spherical void can
indeed form in astrophysical flow systems under the joint action of
thermal pressure force and self-gravity. We expect that such
processes could happen in association with supernova explosions and
evolution of supernova remnants.
%Since the effect of expansion of the universe is crucial in large scale,
%a possible interpretation of our results is that in a relatively small
%scale where the expansion effect is not important, a void can also be
%constructed due the competition between the pressure outwards and
%self-gravity inwards. {\bf Precise meaning? references?}

At the beginning of our model formulation, several physical effects,
such as nuclear reactions, radiation pressure, neutrino transport,
general relativistic effects, rotational effects and magnetic field,
are not taken into account. Under various situations, these effects
can be very important in real astrophysical systems. Given these
approximations and idealizations of our model, it is still hoped
that this simple theoretical model framework may catch certain
essential characteristics or features of flow phenomena of relevant
scenarios and interpretations.
%We may try to include a random magnetic field into this
%model consideration for a further theoretical development.

\section*{Acknowledgments}
This research has been supported in part
%by the Special Funds for Major State Basic
%Science Research Projects of China,
%by the Tsinghua Center for Astrophysics,
%by the Collaborative Research Fund from the National Science
%Foundation of China (NSFC) for Young Outstanding Overseas
%Chinese Scholars (NSFC 10028306) at the National Astronomical
%Observatories, Chinese Academy of Sciences,
by the National Natural Science Foundation of China (NSFC) grants
10373009 and 10533020 at Tsinghua University, by the SRFDP
20050003088, the Yangtze Endowment and the National Undergraduate
Innovation Training Project from the Ministry of Education at
Tsinghua University, by Tsinghua Center for Astrophysics (THCA),
and by the ASCI Center for Astrophysical Thermonuclear Flashes at
the University of Chicago.
%Affiliated institutions of Y-Q

\bsp \vskip 1.0cm

%{\bf Please follow the standard style of reference and take care
%of the details. }

\begin{appendix}

\section{The explicit form of $v'$ and $\alpha'$}\label{a1}

Using Cramer's rule in equations (\ref{eq11}) and (\ref{eq12}), one
can easily deduce the explicit forms of $v'(x)$ and $\alpha'(x)$,
namely,
\begin{eqnarray}
v'(x)={\cal V}(x)/{\cal D}(x)\ ,\\
\alpha'(x)/\alpha(x)={\cal A}(x)/{\cal D}(x)\ ,
\end{eqnarray}
where
\begin{eqnarray}
{\cal V}(x)\equiv -\frac{(ax+v)^2}{(3a+2)}\alpha
+(a+1)(ax+v)v\nonumber\qquad\qquad\qquad\\
-\left(6+3a-\frac{4v}{x}\right)\frac{2C_0}{3}
\left(\frac{ax+v}{3a+2}\right)^{2/3}x^{4/3}\alpha\ ,\quad\\
{\cal A}(x)\equiv2(ax+v)\left(1-\frac{v}{x}\right)
+\frac{(ax+v)}{(3a+2)}\alpha-(a+1)v\qquad\nonumber\\
+\frac{2C_0}{3}\left(\frac{ax+v}{3a+2}\right)^{-1/3}
x^{4/3}\alpha\ ,\qquad\\
{\cal D}(x)\equiv (ax+v)^2-\frac{4C_0}{3}x^{4/3}
\left(\frac{ax+v}{3a+2}\right)^{2/3}\alpha\ .\qquad\qquad
\end{eqnarray}
The sonic singular surface corresponds to ${\cal D}(x)=0$; together
with either ${\cal V}(x)=0$ or ${\cal A}(x)=0$, the sonic critical
line is then determined. Note that ${\cal V}(x)=0$ is equivalent to
${\cal A}(x)=0$ on the sonic singular surface.

\section{Existence condition for
Larson-Penston type solutions}\label{LP}

Starting from equations (\ref{eq1})$-$(\ref{eq5}), we briefly
discuss the existence of Larson-Penston (LP) type solutions for a
general case of $a\neq-2/3$ without constraining $\gamma$. In
fact, equation (\ref{eq5}) is equivalent to equations (\ref{eq1})
and (\ref{eq2}). Assuming Taylor series expansions in the vicinity
of $x=0$, we write the solutions as
\begin{eqnarray}
v(x)=\sum_{k=0}^{\infty}v_kx^k\
,\qquad\qquad\alpha(x)=\sum_{l=0}^{\infty}\alpha_lx^l\ ,
\end{eqnarray}
where $v_k$ and $\alpha_l$ are constant coefficients with
$\alpha_0\neq 0$. By equation (\ref{eq2}), the enclosed mass is
given by
\begin{eqnarray}
m(x)=\int_0^xy^2\alpha(y)dy=\sum_{l=0}^{\infty}
\frac{\alpha_l}{(l+3)}x^{l+3}\ .
\end{eqnarray}
While from equations (\ref{eq1}) and (\ref{eq2}), we have
\begin{eqnarray}
m(x)&=&\frac{(ax+v)}{(3a+2)}x^2\alpha\nonumber\\
 &=&\frac{x^2}{(3a+2)}\left(ax+\sum_{k=0}^{\infty}
 v_kx^k\right)\left(\sum_{l=0}^{\infty}\alpha_lx^l\right)\ .
\end{eqnarray}
The two expressions of $m(x)$ should be equal, giving rise to a
series of relations among the coefficients of $\alpha(x)$ and
$v(x)$,
\begin{eqnarray}
v_0=0\ ,\qquad v_1=2/3\qquad\qquad\qquad\\
4(a+v_1)\alpha_1+4v_2\alpha_0=(3a+2)\alpha_1\\
\vdots\nonumber
\end{eqnarray}
Besides, the reduced pressure can also be written as a series
expansion in the form of
\begin{eqnarray}
p=C_0m^q\alpha^\gamma=C_0x^{3q}
\left[\sum_{l=0}^{\infty}\frac{\alpha_l} {(l+3)}x^{l}\right]^q
\left(\sum_{k=0}^{\infty}\alpha_kx^k\right)^\gamma\ .\label{aeq1}
\end{eqnarray}
Substituting all these series expansions into equation (\ref{eq3})
and comparing coefficients of the same powers of $x$, we have the
following conclusions. For an arbitrary $q$ in general, consider
the power factor $x^{3q}$ in equation (\ref{aeq1}). If $3q$ is not
an integer, the power index of every term in $p$ is not an
integer, and the terms thus cannot have the same power of other
terms in equation (\ref{eq3}). Consequently, no such a series
solution exists. A necessary condition for the existence of LP
type solution is that $q$ takes the form of $J/3$ with $J$ being
an integer. For example, $J=0,\ q=0$ and thus $a=\gamma-2$ and
$C_0=1$, we can readily obtain the following asymptotic solution
(equations 28a and 28b in Suto \& Silk 1988), namely
\begin{eqnarray}
v(x)=\frac{2}{3}x+\frac{\alpha_0^{1-\gamma}}{15\gamma}
\left(\alpha_0-\frac{2}{3}\right)
\left(a+\frac{2}{3}\right)x^3+\cdots\ ,\\
\alpha(x)=\alpha_0-\frac{\alpha_0^{2-\gamma}}{6\gamma}
\left(\alpha_0-\frac{2}{3}\right)x^2+\cdots\ .\qquad\qquad
\end{eqnarray}

For $J=2,\ q=2/3$ and $p=C_0m^{2/3}\alpha^{4/3}$,
%{\bf You mean $\alpha$ here?},
one readily obtains
\begin{eqnarray}
v_0=0\ ,\qquad\quad v_1=2/3\ ,\quad\qquad v_2=0\ ,\qquad\qquad\quad\\
\alpha_0=\frac{2}{3}\left(1+2\sqrt[3]{3}C_0\right)^{-1}\ ,\qquad
\alpha_1=0\ .\qquad\qquad\qquad
\end{eqnarray}
%{\bf I added $v_0=0$ and $v_1=2/3$ here.}
For any given index integer $k>1$ in the series expansion, if we
have already determined coefficients $\alpha_i$ and $v_{i+1}$ where
$0\leq i \leq k-1$, a comparison of the coefficients of each side of
ODEs (\ref{eq1}) and (\ref{eq3}) will give a pair of linear
equations for $\alpha_k$ and $v_{k+1}$, which has a unique solution
for $\alpha_k$ and $v_{k+1}$. Thus, coefficients $\alpha_i$ and
$v_i$ ($i\geq1$) have only one solution. On the other hand, $v_i=0$
and $\alpha_i=0$ ($i\geq1$) gives a solution to this problem.
Consequently for $q=2/3$, only the Einstein-de Sitter solution
exists and no LP type of solution can be found. One possible yet
rare exception occurs when the coefficient determinant of one of the
linear equations for $v_i$ and $\alpha_i$ becomes zero. We do not
give more calculations on these special cases in this paper.

%{\it
\section{Requirement on $C_0$ for the existence of asymptotic
solution (59)}
%(\ref{calculateR})}

We denote the left-hand side of equation (\ref{calculateR}) by
$h(R)$ where $R<2/3$ is required by $\lambda<0$. For a sufficiently
large value of $|R|$ with $R<0$, we have $h(R)>0$; and for
$R\rightarrow2/3^-$, we also have a positive $h(R)$. Taking the
first derivative of $h(R)$ and setting it equal to $0$, we obtain
only one root denoted by $R_0$, namely
\begin{eqnarray}
R_0=(6C_0)^{3/4}(3a+2)-a
\end{eqnarray}
for the minimum of $h(R)$.
%{\bf should check $3/4$.} {\bf It is correct as confirmed by Cao Yi.}
Therefore if $h(R_0)$ is also larger than $0$, then equation
$h(R)=0$ has no real roots and hence asymptotic solution
(\ref{calculateR}) does not exist. On the other hand, if $h(R_0)$ is
smaller than zero, then equation $h(R)=0$ always has two real roots.
The critical case of $h(R_0)=0$ corresponds to a double root, and a
critical value of $C_0=2^{4/3}/6\approx 0.4200$ is thus known.
%}

\end{appendix}

\end{document}